\documentclass[12pt]{article}
\usepackage{epsfig}
\usepackage{psfrag}
\usepackage{latexsym}
\def\beq{\begin{equation}}
\def\eeq{\end{equation}}
\def\bea{\begin{eqnarray}}
\def\eea{\end{eqnarray}}
\def\bet{\begin{tabular}}
\def\eet{\end{tabular}}
\def\bes{\begin{subequations}\bea}
\def\ees{\eea\end{subequations}}

\def\b{\beta}

\newcommand{\rbar}[1]{\overline{#1}}
\def\be{\begin{equation}}
\def\ee{\end{equation}}
\def\bc{\begin{center}}
\def\ec{\end{center}}
\def\bea{\begin{eqnarray}}
\def\eea{\end{eqnarray}}
\def\dd{\displaystyle}
\def\nn{\nonumber}

\catcode`@=11
\def\marginnote#1{}
\newcount\hour
\newcount\minute
\newtoks\amorpm
\hour=\time\divide\hour by60
\minute=\time{\multiply\hour by60 \global\advance\minute by-\hour}
\edef\standardtime{{\ifnum\hour<12 \global\amorpm={am}%
        \else\global\amorpm={pm}\advance\hour by-12 \fi
        \ifnum\hour=0 \hour=12 \fi
        \number\hour:\ifnum\minute<10 0\fi\number\minute\the\amorpm}}
\edef\militarytime{\number\hour:\ifnum\minute<10 0\fi\number\minute}
\def\draftlabel#1{{\@bsphack\if@filesw {\let\thepage\relax
   \xdef\@gtempa{\write\@auxout{\string
      \newlabel{#1}{{\@currentlabel}{\thepage}}}}}\@gtempa
   \if@nobreak \ifvmode\nobreak\fi\fi\fi\@esphack}
        \gdef\@eqnlabel{#1}}
\def\@eqnlabel{}
\def\@vacuum{}
\def\draftmarginnote#1{\marginpar{\raggedright\scriptsize\tt#1}}
\def\draft{\oddsidemargin 0.0truein
        \def\@oddfoot{\sl preliminary draft \hfil
        \rm\thepage\hfil\sl\today\quad\militarytime}
        \let\@evenfoot\@oddfoot \overfullrule 3pt
        \let\label=\draftlabel
        \let\marginnote=\draftmarginnote
   \def\@eqnnum{(\theequation)\rlap{\kern\marginparsep\tt\@eqnlabel}%
\global\let\@eqnlabel\@vacuum}  }
\catcode`@=12
\begin{document}
\begin{titlepage}
\vspace*{-1cm}
\phantom{hep-ph/0501086} 
\hfill{DFPD-05/TH/01}

\vskip 0.5cm
\begin{center}
{\Large\bf Proton Lifetime from SU(5) Unification\\
\vskip .1cm
in Extra Dimensions}
\end{center}
\vskip 0.5  cm
\begin{center}
{\large Maria Laura Alciati}~\footnote{e-mail address: maria.laura.alciati@pd.infn.it}
~~~{\large Ferruccio Feruglio}~\footnote{e-mail address: feruglio@pd.infn.it}
\\
\vskip .1cm
Dipartimento di Fisica `G.~Galilei', Universit\`a di Padova 
\\ 
INFN, Sezione di Padova, Via Marzolo~8, I-35131 Padua, Italy
\\
\vskip .2cm
{\large Yin Lin}~\footnote{e-mail address: y.lin@sns.it}
{\large and}
{\large Alvise Varagnolo}~\footnote{e-mail address: a.varagnolo@sns.it}
\\
\vskip .1cm
Scuola Normale Superiore, Pisa 
\\ 
INFN, Sezione di Pisa, I-56126 Pisa, Italy
\\
\end{center}
\vskip 0.7cm
\begin{abstract}
\noindent
We provide detailed estimates of the proton lifetime in the context of 
simple supersymmetric SU(5) grand unified models
with an extra compact spatial dimension, described by the orbifold
$S^1/(Z_2\times Z_2')$ and by a large compactification scale
$M_c\approx 10^{14}\div 10^{16}$ GeV. 
We focus on a class of models where the grand 
unified symmetry is broken by the compactification mechanism
and where baryon violation proceeds mainly through gauge vector boson 
exchange so that the proton lifetime scales as $M_c^4$.
We carefully compute $M_c$ from a next-to-leading analysis of gauge
coupling unification and we find that $M_c$ can only be predicted
up to an overall factor $10^{\pm 1}$. 
The simplest model, where the dominant decay mode is $\pi^0 e^+$ and 
has no flavour suppression, is strongly constrained by existing data,
but not totally ruled out. We also analyze models where some of the matter 
fields are localized in the extra space and proton decay is flavour 
suppressed. In models associated to anarchy in the neutrino sector
the preferred decay channel is $K^+ {\bar \nu}$ and the lifetime can be 
within the reach of the next generation of experiments.
\end{abstract}
\end{titlepage}
\setcounter{footnote}{0}
\vskip2truecm

%
%%%%%%%%%%%%%%%%%%%%%%%    1. INTRODUCTION       %%%%%%%%%%%%%%%%%%%%%
%
\section{Introduction}
Grand unification can be considered a cornerstone in our quest
for unification of particle interactions. Many properties of the standard
model of electroweak and strong interactions that seem mysterious
or accidental, like the particle content, the cancellation of gauge
anomalies, the quantization of the electric charge, appear natural
in the context of grand unified theories (GUTs). The quantitative success
of gauge coupling unification in supersymmetric (SUSY) GUTs is remarkable
and the vicinity of the GUT scale to the Planck scale is quite
intriguing.
Also the observed smallness of neutrino masses, suggesting
the breaking of the total lepton number L at very high energies,
supports the idea of a new threshold in particle physics
at the GUT scale. 

Despite the attractiveness of the GUT idea and its influence in both
theoretical and experimental developments, proton decay, that represents
its most characteristic prediction, has eluded all the experimental
searches so far. Stringent limits on the proton lifetime in many 
channels have been set, such as \cite{shio}
\bea
\tau(p\to \pi^0 e^+)&>& 5.4\times 10^{33}~{\rm yr}~~~~~~~ 90\% {\rm C.L.}\nn\\
\tau(p\to K^+ {\bar\nu})&>& 2.2\times 10^{33}~{\rm yr}~~~~~~~ 90\% {\rm C.L.} 
\label{expbound}
\eea
and, as a consequence, the minimal SUSY GUTs are 
tightly constrained 
or even ruled out. The proton decay rate in minimal SUSY GUTs
depends on many parameters and there are many sources of theoretical
uncertainties, such as the mass of the colour triplets whose
exchange dominates the amplitudes, the hadronic 
matrix elements, the spectrum of SUSY particles, $\tan\beta$, 
unknown phases and mixing angles from the flavour sector. 
Even by stretching the uncertainties to their limits, minimal SUSY SU(5)
is already ruled out by now \cite{minimal} and, in general, only non-minimal 
schemes, considerably more complicated than the minimal ones, 
survive this important test \cite{nonminimal}. On the theoretical side
a related difficulty of minimal GUTs
is how to reconcile light electroweak doublets with
superheavy colour triplets, both occurring in the same GUT
multiplets: the so-called doublet-triplet splitting problem.
In minimal GUTs such a splitting is achieved through a fine-tuning
by fourteen orders of magnitude.
Even when such a splitting is obtained, either by fine-tuning or by another
mechanism, in many models it is upset by radiative corrections
when SUSY is broken and/or by non-renormalizable operators which can
originate from the physics at the cut-off scale \cite{dt}.
 
An appealing mechanism to achieve the desired splitting is 
when the grand unified symmetry is broken by the compactification 
mechanism in models with extra spatial dimensions \cite{w}.
In the last years this mechanism has been reproduced in the
context of simple field theoretical models, GUTs defined
in five or more space-time dimensions \cite{kawa}. In the simplest case 
a single extra spatial dimension is compactified on a circle.
The Lagrangian of the theory is invariant under SU(5), but the fields 
are periodic only up to transformations leaving the SU(5) algebra invariant. 
This requirement
produces automatically a mass splitting in complete SU(5) multiplet.
For instance, in the gauge vector boson sector, the only fields
that remain massless correspond to the vector bosons of the 
standard model. All the other states, including
the extra gauge bosons $X$ of SU(5), become massive, with masses of the
order of the compactification scale $M_c\equiv 1/R$. 
This is a simple and elegant 
way to effectively break SU(5) down to the standard model gauge group.
The same boundary conditions that break the grand unified symmetry
should be consistently extended to the other fields of the theory
and in particular to the multiplet containing the electroweak doublets
and colour triplets. As a consequence, an automatic doublet-triplet splitting
of order $1/R$ occurs, with no need of an ad hoc scalar sector. 

It has soon been realized that such a framework has interesting 
consequences also on the prediction of proton lifetime. Proton could even 
be made stable in such a construction \cite{af,hebecker}. 
In a less radical approach,
the automatic doublet-triplet splitting is accompanied by
the absence of baryon-violating dimension five operators \cite{hn1}.
Therefore proton decay can only proceed through dimension six
operators, originating from the exchange of the superheavy gauge vector 
bosons $X$ between the ordinary fermions.
In conventional, four-dimensional models, the inverse decay rate
due to dimension six operators is given by \cite{hisano}
\be
\tau(p\to \pi^0 e^+)=1.6\times 10^{36}~{\rm yr}
\times \left(\frac{\alpha~({\rm GeV}^3)}{0.015}\right)^{-2}
\times \left(\frac{\alpha_U}{1/25}\right)^{-2}
\times \left(\frac{M_X~({\rm GeV})}{2\times 10^{16}}\right)^{4}~~~,
\label{4Dmodels}
\ee
where $M_X$ is the mass of the gauge vector bosons $X$, 
$\alpha_U$ is the gauge coupling constant at the unification 
scale and $\alpha$ is a parameter coming from the evaluation of the 
hadronic matrix element. The central value in (\ref{4Dmodels}) is too
large to be observed, even by the next generation of experiments.
In five dimensional models there are important modifications 
and the prediction, even in the most conservative case, can be quite
different from the one in (\ref{4Dmodels}). First the proton lifetime
scales with the fourth power of the compactification scale $M_c$,
which can be considerably smaller than the central value of $M_X$
in eq. (\ref{4Dmodels}). Second, the coupling of the gauge vector 
bosons $X$ to ordinary fermions differs, in general, from  
four-dimensional SUSY SU(5). This has to do with the freedom of
introducing matter fields either as `bulk' fields, depending on the extra 
spatial dimension, or as `brane' fields, localized in a particular
four-dimensional subspace of the full space-time.

In this paper we focus on the five-dimensional SUSY SU(5)
GUT described in ref. \cite{hn1,n2,hn3}, which can be considered among the
simplest realizations of GUTs in extra dimensions
\footnote{Proton decay in other GUTs with extra dimensions has  
been analyzed in \cite{oth}.}. 
There are estimates of the proton decay rate in this model and its
variants. In ref. \cite{hn1,n2} the compactification scale $M_c$ has been
evaluated from the analysis of gauge coupling unification in a leading 
order approximation. By calling $\Lambda$ the cut-off scale of the theory,
it has been shown that, for $\Lambda R$ equal to 100 (10), the 
compactification scale ranges in
\be
5\times 10^{14}~(3\times 10^{15})~{\rm GeV}\le M_c\le 
2\times 10^{15}~(8\times 10^{15})~{\rm GeV}~~~.
\label{mcllog}
\ee
Indeed $\Lambda R\approx 100$ is preferred by a more refined estimate
\cite{cprt}
and, if we naively substitute $2 M_c/\pi$ to $M_X$ in eq. (\ref{4Dmodels}),
we find a proton lifetime
\be
1.0\times 10^{29}~{\rm yr}\le\tau(p\to \pi^0 e^+)\le 2.6\times 10^{31}~{\rm yr}~~~,
\ee
already excluded by the experimental bound of eq. (\ref{expbound}).
In this first estimate, however, it has been assumed that all matter fields
are coupled to the $X$ bosons as in the four-dimensional theory.
Actually, a more suitable setting for a natural interpretation of the 
fermion mass 
hierarchies requires that only $T_3$ and $F_i$ $(i=1,2,3)$
couple to the $X$ boson as in the four-dimensional theory \cite{n2,hn3,hmr}.
Here $T$ and $F$ denote 10 and ${\bar 5}$ SU(5) representations and
the index labels the different generations.
In such a framework the leading baryon-violating dimension six operator 
is $T^\dagger_3 T_3 F^\dagger_1 F_1$. Mixing angles are required to 
convert third generation fields into states relevant to proton decay
and the amplitude for the dominant decay mode, $K^+ {\bar \nu}$, 
is depleted by $\lambda^5$ with respect to $\pi^0 e^+$
in the four-dimensional theory, $\lambda\approx 0.22$ being a Cabibbo
suppression factor. The conclusion of ref. \cite{hn3} is that
\be
\tau(p\to K^+ {\bar \nu})\approx 10^{37\pm 2}~{\rm yr}~~,
\label{hnpl}
\ee
unless additional, model dependent contributions are invoked.

The result (\ref{hnpl}) is not far from the expected sensitivity
of future Water Cerenkov and Liquid Argon detectors for the 
$K^+ {\bar \nu}$ mode \cite{futsens}:
\bea
0.2\times 10^{35}~{\rm yr}&~&~~~~~~~~~ {\rm 10 yr, 650 kton}~~~~~~~~~~{\tt Water~ Cherenkov}\nn\\
1.1\times 10^{35}~{\rm yr}&~&~~~~~~~~~ {\rm 10 yr, 100 kton}~~~~~~~~~~{\tt Liquid~ Argon}
\eea
This strongly motivates a careful and complete quantitative estimate
of the proton lifetime in this model, with a 
particular attention to the main sources of theoretical 
uncertainties, that can sizeably affect the estimate in eq. (\ref{hnpl}).
Such an analysis is the purpose of this paper.
First of all we compute $M_c$ from a next-to-leading analysis of gauge coupling
unification, including two-loop running, threshold corrections
from SUSY particles,
and possible SU(5) breaking contributions allowed by the present
construction. As we will see in detail in section 4, we find that
there is a big theoretical uncertainty on $M_c$, entirely
dominated by the expected SU(5)-breaking contributions and not 
accounted for by the previous estimates. According to our analysis
$M_c$ can range from approximately $10^{14}~{\rm GeV}$ to more than
$10^{16}~{\rm GeV}$, a considerably wider interval than in eq. 
(\ref{mcllog}). We also size the uncertainty coming from the flavour sector, 
by contemplating different settings for matter fields.
In what we call `option 0', all matter fields are coupled to the $X$
gauge bosons as in four-dimensional SUSY SU(5). 
We will also consider two other possibilities,
that we call `option I' and `option II'. Option II corresponds to
$T_3$ and $F_i$ $(i=1,2,3)$ coupled to $X$ as in four dimensions, whereas
in option I only $T_3$ and $F_{3,2}$ have standard couplings.
As far as neutrino mass matrices are concerned, option I and option II
give rise to ``semianarchy'' and ``anarchy'', respectively \cite{review}.
We estimate the degree of flavour suppression by counting
powers of the Cabibbo angle $\lambda$, in two different
ways, characterized by a stronger and a milder mixing between
the different generations. This allows us to estimate also the theoretical
uncertainty coming from the Yukawa sector.

Our results are presented in section 5.
They are qualitatively different for option 0 
and options I/II.
In option 0, the dominant decay mode
is $\pi^0 e^+$ and the lifetime is obtained from
eq. (\ref{4Dmodels}) through the substitution $M_X\to 2 M_c/\pi$
\footnote{The numerical factor $2/\pi$ is due to the exchange of
a whole tower of Kaluza-Klein modes.}.
Due to the large uncertainty on $M_c$ we will see that such an option
is strongly constrained, but not totally ruled out. In option I and II,
we confirm that $K^+ {\bar \nu}$
is one of the dominant decay channels.
The prospects for detectability of proton decay in this channel
by future machines are good, but only 
with favorable combinations of mixing angles (option II is needed)
and of SU(5)-violating terms at the cut-off scale.
We also evaluate branching ratios for proton decay,
much less affected by uncertainties than absolute rates.

%
%%%%%%%%%%%%%%%%%%%%%%%    2. THE MODEL       %%%%%%%%%%%%%%%%%%%%%
%
\section{SU(5) unification in extra dimensions}

\subsection{Space-time orbifold}

Following ref. \cite{kawa} we consider a 5-dimensional space-time factorized into a product of the ordinary 
4-dimensional space-time M$_4$ and of the orbifold $S^1/(Z_2\times Z_2')$, with coordinates $x^{\mu}$, 
($\mu=0,1,2,3$) and $y=x^5$. The orbifold $S^1/(Z_2\times Z_2')$ 
is obtained from the circle $S^1$ of radius $R$ 
with the identification provided by the two reflections:
\be
Z_2:~~~~~~~~~~y\rightarrow -y~~~~~,
\ee
\be
Z_2':~~~~~~~~~y'\rightarrow -y'~~~~~(y'\equiv y-\pi R/2)~~~~~.
\ee
As a fundamental region of the orbifold, 
we can take the interval from $y=0$ to $y=\pi R/2$. 
At the two sides of the interval,
we have two four-dimensional boundaries of the space-time,
called branes. On the covering space, $S^1$, 
where we choose to work from now on, the origin $y=0$
and $y=\pi R$ represent the same physical point and similarly for
$y=+\pi R/2$ and $y=-\pi R/2$. When speaking of the brane at 
$y=0$, we actually mean the two four-dimensional slices at  
$y=0$ and $y=\pi R$, and similarly $y=\pi R/2$ stands for both $y=\pm\pi R/2$.

For a generic field
$\phi(x^{\mu},y)$ living in the 5-dimensional bulk the $Z_2$ and $Z_2'$ parities $P$ and $P'$ are defined by 
\bea
\phi(x^{\mu},y)&\to \phi(x^{\mu},-y)=P\phi(x^{\mu},y)~~~,\nn\\
\phi(x^{\mu},y')&\to \phi(x^{\mu},-y')=P'\phi(x^{\mu},y')~~~.
\eea
Denoting by $\phi_{\pm \pm}$ the fields with $(P,P')=(\pm,\pm)$ we have the $y$-Fourier expansions:
\bea
  \phi_{++} (x^\mu, y) &=& 
\sqrt{1 \over {2\pi R}}\phi^{(0)}_{++}(x^\mu)+
       \sqrt{1 \over {\pi R}} 
      \sum_{n=1}^{\infty} \phi^{(2n)}_{++}(x^\mu) \cos{2n y \over R}~~~,
\label{phi++exp}\nn\\
  \phi_{+-} (x^\mu, y) &=& 
       \sqrt{1 \over {\pi R}} 
      \sum_{n=0}^{\infty} \phi^{(2n+1)}_{+-}(x^\mu) \cos{(2n+1)y \over R}~~~,
\label{phi+-exp}\nn\\
  \phi_{-+} (x^\mu, y) &=& 
       \sqrt{1 \over {\pi R}}
      \sum_{n=0}^{\infty} \phi^{(2n+1)}_{-+}(x^\mu) \sin{(2n+1)y \over R}~~~,
\label{phi-+exp}\nn\\
 \phi_{--} (x^\mu, y) &=& 
       \sqrt{1 \over {\pi R}}
      \sum_{n=0}^{\infty} \phi^{(2n+2)}_{--}(x^\mu) \sin{(2n+2)y \over R}~~~.
\label{fourier}
\eea
where $n$ is a non negative integer and the notation is such that
the Fourier component field $\phi^{(n)}(x)$
acquires a mass $n/R$ upon compactification.
Only $\phi_{++}$ has a massless component and only $\phi_{++}$ and $\phi_{+-}$ are non-vanishing on the 
$y=0$ brane. The fields $\phi_{++}$ and $\phi_{-+}$ are non-vanishing on the $y=\pi R/2$ brane, 
while $\phi_{--}$ vanishes on both branes. The normalization is such that:
\be
\int_{-\pi R}^{+\pi R} dy \vert\phi(x,y)\vert^2=\sum_{n=0}^\infty \vert\phi^{(n)}(x)\vert^2~~~.
\ee

\subsection{Gauge and Higgs sectors}

The theory under investigation, defined in the above five-dimensional
space-time, is invariant under N=1 SUSY, which corresponds
to N=2 in four dimensions, and under SU(5) gauge transformations. We can keep working in the entire circle
if we also ask invariance under $Z_2\times Z_2'$ parities.  
The parities of the fields are assigned in such a way that compactification reduces 
N=2 to N=1 SUSY and breaks SU(5) down to the SM  gauge group ${\rm SU(3)} \times {\rm SU(2)} \times
{\rm U(1)}$. Leaving aside for the moment quarks and leptons and their SUSY partners, 
the 5-dimensional theory contains the following N=2 SUSY multiplets of fields. 
First, a vector multiplet with vector bosons $A_M$, $M=0,1,2,3,5$, two bispinors $\lambda^i$, 
$i=1,2$, and a real scalar $\Sigma$, each of them transforming as a 24 representation of SU(5).
\\[0.1cm]
\begin{table}[h]
{\begin{center}
\begin{tabular}{|c|c|c|}   
\hline
& & \\                         
$(P,P')$ & field & mass\\ 
& & \\
\hline
& & \\
$(+,+)$ &  $A^a_{\mu}$, $\lambda^{2a}$, $H^D_u$, $H^D_d$ & $\frac{2n}{R}$\\
& & \\
\hline
& & \\
$(+,-)$ &  $A^{\hat{a}}_{\mu}$, $\lambda^{2\hat{a}}$, $H^T_u$, $H^T_d$ & $\frac{(2n+1)}{R}$ \\
& & \\ 
\hline
& & \\
$(-,+)$ &  $A^{\hat{a}}_5$, $\Sigma^{\hat{a}}$, $\lambda^{1\hat{a}}$, $\hat{H}^T_u$, $\hat{H}^T_d$  
& $\frac{(2n+1)}{R}$\\
& & \\
\hline
& & \\
$(-,-)$ &  $A^{a}_5$, $\Sigma^{a}$, $\lambda^{1a}$, $\hat{H}^D_u$, $\hat{H}^D_d$ & $\frac{(2n+2)}{R}$ \\
& & \\
\hline
\end{tabular} 
\end{center}}
\caption{Parity assignment and masses ($n\ge 0$) of fields in the vector and Higgs supermultiplets.}
\label{t1}
\end{table}
\\[0.3cm]
These fields can be arranged in an N=1 vector supermultiplet $V\equiv(A_\mu,\lambda^2)$ and 
an N=1 chiral multiplet $\Phi\equiv(A_5,\Sigma,\lambda^1)$, both transforming in the adjoint 
representation of SU(5).
Then there are two hypermultiplets $H^s$ ($s=1,2$), each of them
consisting of two N=1 chiral multiplets: $H^1=(H_5,\hat{H}_{\bar 5})$
and $H^2=(H_{\bar 5},\hat{H}_5)$.
Here $H_5$ and $H_{\bar 5}$ stand for the usual SU(5) supermultiplets, 
which include the scalar Higgs doublets $H^D_u$ and $H^D_d$ and the 
corresponding scalar triplets $H^T_u$ and $H^T_d$, respectively. 
The other two, $\hat{H}_{\bar 5}$ and 
$\hat{H}_5$, transform as $\bar{5}$ and 5, under SU(5).
The parity $P$, $P'$ assignments are the same as in ref. \cite{kawa} and are given in table 1.

Here the index $a$ ($\hat{a}$) labels the unbroken (broken) SU(5) generators $T^a$ ($T^{\hat{a}}$), 
$H$ stands for the whole chiral multiplet of given quantum numbers. 
The parity $P$ breaks N=2 SUSY down to N=1 and would allow complete SU(5) massless supermultiplets,
contained in the first two rows of table 1. The additional parity $P'$ respects the surviving N=1 SUSY and breaks
SU(5) down to the standard model gauge group. Note that the derivative $\partial_5$ 
transforms as $(-,-)$. The $(+,+)$ fields, which remain
massless and do not vanish on both branes are the gauge and Higgs multiplets of the low 
energy Minimal SUSY Model (MSSM). The bulk 5-dimensional Lagrangian is exactly as in ref. \cite{kawa} and we do not 
reproduce it here. 

The vector bosons $A_\mu^{a(0)}$
remain massless and, together with the gauginos 
$\lambda^{2a(0)}$, form a vector supermultiplet of N=1. 
All the other vector bosons are massive and become component of 
N=1 massive vector supermultiplets. Therefore the SU(5) gauge symmetry
is effectively broken down to SU(3) $\times$ SU(2) $\times$ U(1)
and the symmetry breaking order parameter is the inverse radius
$1/R$, which is expected to be of the order of the unification
scale.
At the same time, in the Higgs sector, only the Higgs doublets
and their superpartners are massless, while colour triplets and
extra states acquire masses of order $1/R$, giving rise to
an automatic doublet-triplet splitting. The N=2 SUSY algebra would
allow a mass term between the hypermultiplets $H^1$ and $H^2$  
\cite{sohnius}, thus spoiling the lightness of the Higgs doublets achieved by compactification,
but such a term can be forbidden by explicitly requiring an additional
U(1)$_R$ symmetry \cite{hn1,hn3}. 
This U(1)$_R$ symmetry is the diagonal subgroup
of an $R$-symmetry acting on the Higgs multiplets as
\bea
(H_5,\hat{H}_{\bar 5})'(\theta)&=&e^{i\alpha}(H_5,\hat{H}_{\bar 5})(e^{-i\alpha}\theta)\\
(H_{\bar 5},\hat{H}_{5})'(\theta)&=&e^{i\alpha}(H_{\bar 5},\hat{H}_{5})(e^{-i\alpha}\theta)~~~
\eea
and a U(1) symmetry acting as:
\bea
(H_5,{\hat{H}_{\bar 5}}^{\dagger})'&=&e^{-i\alpha}(H_5,{\hat{H}_{\bar 5}}^\dagger)\\
(H_{\bar 5},{\hat{H}_{5}}^{\dagger})'&=&e^{-i\alpha}(H_{\bar 5},{\hat{H}_{5}}^\dagger)
\eea
where $\theta$ denotes the Grassmann coordinate of the N=1 superspace.
The overall $R$ charges of $H_5$, $\hat{H}_{\bar 5}$, $H_{\bar 5}$ and $\hat{H}_{5}$
are 0, 2, 0, 2 respectively.
This symmetry forbids a mass term between $H_5$ and $H_{\bar 5}$.
Therefore, before the breaking of the residual N=1
SUSY, the mass spectrum is the one shown in table 1.

\subsection{Matter fields}

Different options exist to introduce quarks and leptons.
A first possibility is to describe them as N=2 bulk hypermultiplets 
\cite{hebecker}.
In each hypermultiplet the different SU(5) components undergo 
a splitting similar to the one occurring in the gauge and in the Higgs 
sector. Consider for instance a hypermultiplet transforming as
${\bar 5}\oplus 5$ under SU(5), where, in an obvious notation,
${\bar 5}\equiv (L,D^c)$ and $5\equiv (\hat{L},\hat{D^c})$. 
We can assign orbifold parities in such a way that either $L$ or $D^c$
but not both of them contains a fermion zero mode and, to  
get all the massless degrees of freedom of a ${\bar 5}$ representation,
we need to introduce an additional hypermultiplet, 
$(L,D^c)'\oplus (\hat{L},\hat{D^c})'$ with opposite
$P'$ assignment. A similar splitting occurs in 10 representations.
It is also interesting to note that in the bulk Lagrangian, that
conserves the momentum along the fifth dimension, the gauge vector bosons
$A^{\hat{a}}_\mu$ that violate baryon number, never couple two
fermion zero modes of an hypermultiplet. Therefore fermions belonging
to hypermultiplets cannot contribute to proton decay, at least in a 
minimal version of the model \footnote{It is possible to introduce non-diagonal
kinetic terms between matter hypermutiplets $(M,M')$ on the $y=0$ brane
\cite{hn1,hn3}.
Such terms can contain interaction terms between matter zero modes and 
superheavy gauge bosons, thus contributing to proton decay. In our analysis we 
neglect such a possibility that enhances proton decay rates
at the price of introducing a strong model dependence.}.

A second possibility is to identify matter as N=1 SUSY chiral multiplets,
localized at one of the two branes. On the brane at $y=\pi R/2$,
the gauge invariance is limited to the SM gauge transformations
and incomplete SU(5) multiplets can be introduced. 
The kinetic terms of such multiplets, localized at $y=\pi R/2$,
do not contain any baryon or lepton violating interactions.
Therefore such interactions are only possible if we introduce
ad hoc non-minimal terms of the gauge vector bosons $A^{\hat{a}}_\mu$
in $y=\pi R/2$. Although technically possible, we will disregard
such a possibility in the present analysis, not to introduce
too many unknown parameters.
On the brane at $y=0$, where the full SU(5)
gauge invariance is at work, N=1 chiral multiplets should be introduced
as complete SU(5) representations. Moreover, the gauge invariant
kinetic terms defined on the brane do not conserve
the momentum along the extra dimensions, and the fermions are coupled 
to all the gauge vector bosons, including those violating baryon and 
lepton numbers.
In summary, in a minimal model, matter is introduced either in
bulk hypermultiplets or in N=1 chiral multiplets at $y=0$.
Baryon and lepton violation between massless fermions mediated by gauge 
vector bosons is only due to minimal couplings at $y=0$.

Additional sources of baryon and lepton violations can be provided
by Yukawa couplings that, in the present model, are all localized
in $y=0$. Among these couplings there are dimension three and four
operators, which in MSSM are forbidden by R-parity.
The role of R-parity is played in the present context by the U(1)$_R$
R-symmetry discussed in the previous subsection, forbidding a mass term between the Higgs
hypermultiplets $H^1$ and $H^2$. 
As shown in ref. \cite{hn1} and described in table 2,
such a symmetry can be suitably extended to the matter sector 
in such a way that the leading order allowed Yukawa interactions are:
\be
S_Y=\int d^4x dy \delta(y)\int d^2\theta 
\left(T y_u T H_5 + F y_d T H_{\bar 5}+\frac{F H_5 w F H_5}{M}\right) + h.c. 
\label{yuk}
\ee
Here $T$ and $F$ denote 10 and ${\bar 5}$ representations, either
localized in $y=0$ or belonging to bulk hypermultiplets and 
$y_{u,d}$ and $w$ are the Yukawa couplings, 
3$\times$3 matrices in flavour space; $M$ denotes the scale
where the total lepton number is violated. In the limit of exact
residual N=1 SUSY, the U(1)$_R$ symmetry is also exact and 
all the unwanted dimension 3 and 4 operators
are absent. When N=1 SUSY is eventually broken, such operators
are still forbidden if R-parity, contained in U(1)$_R$, survives 
after the breaking.

\begin{table}[h]
\begin{center}
\begin{tabular}[b]{|c|c c|c c|c c|c c|c c|}
\hline
Field & $V$ & $\Sigma$ & $H_5$ & $\hat{H}_{\rbar{5}}$ & $H_{\rbar{5}}$ & $\hat{H}_5$ & $T$ & $\hat{T}$ & $F$ & $\hat{F}$\\
\hline
$U(1)_R$ & $0$ & $0$ & $0$ & $2$ & $0$ & $2$ & $1$ & $1$ & $1$ & $1$\\
\hline
\end{tabular}
\end{center}
\caption{U(1)$_R$ charges for 4D vector and chiral superfields. Matter fields, either in tenplets $T$ or fiveplets $F$ of SU(5) have equal U(1)$_R$ charge, whether they are bulk or brane multiplets.}
\label{table:U1R}
\end{table}

An important consequence for matter stability is that
the dimension 5 operator $TTTF$ is also forbidden
by U(1)$_R$. Indeed, a crucial difference between this model
and minimal SUSY SU(5) is that the colour triplets
whose exchange gives rise to $TTTF$
acquire mass from compactification and not from a term
that mixes $H_5$ and $H_{\bar 5}$. 
When U(1)$_R$ gets broken at low energy
a mixing term
of the order of the SUSY breaking scale arises, but it is
too tiny to produce significant effects.
As a result, the only sizeable source of proton decay in the model
are the dimension 6 operators originating from the exchange 
of the heavy fields $A^{\hat{a}}_\mu$.

The Yukawa interactions of eq. (\ref{yuk}) describe, after
electroweak symmetry breaking, masses for fermions, including
neutrinos. A careful inspection of fermion masses suggests
whether to localize matter multiplets at $y=0$ or to
embed them in bulk hypermultiplets. The most plausible
assignments are reported in table 3. 
\\[0.1cm]
\begin{table}[h]
{\begin{center}
\begin{tabular}{|c||c||c||c|}   
\hline
& & &  \\                         
& {\tt option 0} & {\tt option I}  & {\tt option II} \\ 
& & & \\
\hline
& & & \\                         
{\tt bulk}&  & $T_1$, $T_2$, $F_1$  & $T_1$, $T_2$ \\ 
& & & \\
\hline
& & &  \\                         
{\tt brane} $y=0$& $T_i$, $F_i$ $(i=1,2,3)$& $T_3$, $F_2$, $F_3$  & $T_3$, $F_1$, $F_2$, $F_3$ \\ 
& & & \\
\hline
\end{tabular} 
\end{center}}
\caption{Matter fields and their locations. Indices 1, 2, 3 label
generations. Bulk fields belong to N=2 supermultiplets and 
they should be doubled, to provide the correct number of zero modes.
In the table this doubling is understood. For instance
$T_1$ of option I stands for $T_1$, $T'_1$. Options I and II differ only for 
the assignments of $F_1$.}
\label{t3}
\end{table}
\\[0.3cm]

We shortly recall the origin of this choice. 
Option 0 realizes the simplest five-dimensional realization of SUSY SU(5).
The hierarchy among fermion masses is accommodated by adjusting by hand
the Yukawa couplings. The same fermion mass relations of minimal 
SUSY SU(5) hold.
Option I and II represent alternative description of the flavour sector
that take advantage of the existing freedom in five dimensions.
If both $T_i$ and $F_i$ of the same generation $i$ are put in $y=0$, 
then, as in minimal SUSY SU(5), this implies
the equality between the $ii$ mass matrix elements of 
charged fermions and down quarks.
This is only acceptable, and actually welcome, for the third 
generation. Hence, to begin with, in option I and II, 
$T_3$ and $F_3$ are chosen as brane fields.
On the contrary, if at least one between $T$ and $F$ belong
to an hypermultiplet, the SU(5) relation does not hold any more,
because the masses of charged fermions and down quarks originate
from two different hypermultiplets. 
By assigning $F_2$ to $y=0$ we can naturally explain the largeness
of the 23 mixing in the lepton sector, and, consequently,
$T_2$ is chosen as a bulk field. It is also preferable to
have $T_1$ as a bulk field, which helps in reproducing the
strong hierarchy in the up quark sector. The only ambiguity concerns $F_1$
and we are left with the two possible options, I and II, of table 3.

\subsection{Fermion masses}

Fermion masses in option 0 are introduced by tuning the Yukawa 
couplings $y_u$, $y_d$ and $w$ of eq. (\ref{yuk}), to the desired values.
The resulting description is not completely satisfactory 
due to the equality between down quark and charged lepton masses, 
the well known mass relations of minimal SU(5). These are good for the
third generation but wrong, by factors of order one, for first and second 
generations.
There are several mechanisms to correct the undesired relations 
\cite{ellis} but they
do not appreciably alter the predictions for proton lifetime \cite{mura}.
Therefore, in what follows, we will maintain option 0 in its minimal
form, without introducing any Yukawa couplings other than those
of eq. (\ref{yuk}).

Alternatively, in option I and II, we can take advantage of the fact that
bulk and brane matter fields enter the Yukawa couplings
of eq. (\ref{yuk}) with a different relative normalization
which is induced by the different mass dimension of four
and five dimensional fields. Each Fourier component of a bulk field in eq. (\ref{yuk})  
carries a factor 
\be
\epsilon=\frac{1}{\sqrt{\pi R\Lambda}}
\label{eps}
\ee
as can be seen from the expansion of eq. (\ref{fourier}). 
The mass parameter $\Lambda$, needed for dimensional reason
in the Yukawa couplings, is some cut-off scale of our effective theory.
Notice that $\epsilon$ is always a suppression factor since
the theory only makes sense as an effective theory in the regime 
$\Lambda>1/R$. We can include the appropriate power of $\epsilon$ into the coupling
constants $y_{u,d}$ and $w$. From the assignments of table 3
we can read the relative suppression due to $\epsilon$ 
of the different matrix elements in fermion mass matrices:
\\[0.3cm]
\be
y_u=\left(
\begin{array}{ccc}
\epsilon^2 & \epsilon^2 & \epsilon\\
\epsilon^2 & \epsilon^2 & \epsilon\\
\epsilon & \epsilon & 1
\end{array}
\right)
\label{mu}
\ee
\be
\begin{array}{c}
{\tt option~ I}\\
\\
y_d=\left(
\begin{array}{ccc}
\epsilon^2 & \epsilon^2 & \epsilon\\
\epsilon & \epsilon & 1\\
\epsilon & \epsilon & 1
\end{array}
\right)\\
\\
w=\left(
\begin{array}{ccc}
\epsilon^2 & \epsilon & \epsilon\\
\epsilon & 1 & 1\\
\epsilon & 1 & 1
\end{array}
\right)
\end{array}~~~~~~~~~~~~~~~~~~~~~~~~~
\begin{array}{c}
{\tt option~ II}\\
\\
y_d=\left(
\begin{array}{ccc}
\epsilon & \epsilon & 1\\
\epsilon & \epsilon & 1\\
\epsilon & \epsilon & 1
\end{array}
\right)\\
\\ 
w=\left(
\begin{array}{ccc}
1 & 1 & 1\\
1 & 1 & 1\\
1 & 1 & 1
\end{array}
\right)
\end{array}~~~~~,
\label{denu}
\ee
\\[0.3cm]
where the matrices are given up to overall factors
and up to further order-one dimensionless coefficients that might affect
each individual entry. The charged lepton mass matrix
$m_e$ is approximately given by $m_d^T$, the exact equality
being broken by the first and second generations.
An interesting structure shows up
in the 23 sector. If $\epsilon$ is of order $\lambda^2$,
where $\lambda\approx 0.22$ denotes a Cabibbo suppression factor,
then $m_c/m_t\approx \lambda^4$, $m_s/m_b\approx \lambda^2$,
and $V_{cb}\approx\lambda^2$.
In the lepton sector we have $m_\mu/m_\tau\approx \lambda^2$, 
$\theta_{23}\approx O(1)$
and all this goes into the right direction.
From eq. (\ref{mu},\ref{denu}) we see that the first generation is 
not equally well described. We can adjust first generation
couplings by exploiting the freedom related to the dimensionless
coefficients, which in eq. (\ref{mu},\ref{denu}) have been set to one.   
Alternatively we can improve the description of fermion masses
by slightly modifying the wave function for the zero mode of $T_1$
with respect to that of $T_2$. So far, these wave functions have been taken 
constant in $y$ with the same value at $y=0$.
If the wave function
for the zero mode of $T_1$ is displaced from $y=0$ there will be
a relative suppression between the couplings of $T_1$ and $T_2$.
Such a change in the $y$-profile of the wave function
for the zero mode in $T_1$
can be easily obtained by introducing a 
bulk mass term for the hypermultiplet containing $T_1$ and 
the net effect can be described by replacing $\epsilon$ with a new
suppression factor $\delta<\epsilon$ in the matrix elements 
involving $T_1$ \cite{bulkmass}. We get:  
\\[0.3cm]
\be
y_u=\left(
\begin{array}{ccc}
\delta^2 & \epsilon\delta & \delta\\
\epsilon\delta & \epsilon^2 & \epsilon\\
\delta & \epsilon & 1
\end{array}
\right)
\label{newmu}
\ee
\be
\begin{array}{c}
{\tt option~ I}\\
\\
y_d=\left(
\begin{array}{ccc}
\epsilon\delta & \epsilon^2 & \epsilon\\
\delta & \epsilon & 1\\
\delta & \epsilon & 1
\end{array}
\right)\\
\end{array}~~~~~~~~~~~~~~~~~~~~~~~~~
\begin{array}{c}
{\tt option~ II}\\
\\
y_d=\left(
\begin{array}{ccc}
\delta & \epsilon & 1\\
\delta & \epsilon & 1\\
\delta & \epsilon & 1
\end{array}
\right)\\
\end{array}~~~~~.
\label{newdenu}
\ee
\\[0.3cm]
By assuming that the unknown coefficients multiplying
the different matrix elements are of order one, we get
the order of magnitude relations:
\be
\begin{array}{lcl}
\dd\frac{m_c}{m_t}\approx \epsilon^2 &~~~~~~~~~~ &
\dd\frac{m_u}{m_t}\approx \delta^2\nn\\
\dd\frac{m_s}{m_b}\approx\frac{m_\mu}{m_\tau}\approx \epsilon &~~~~~~~~~~ &
\dd\frac{m_d}{m_b}\approx\frac{m_e}{m_\tau}\approx 
\left\{
\begin{array}{cc}
\epsilon\delta& ({\tt option I})\\
\delta& ({\tt option II})
\end{array}
\right.
\end{array}
\ee
\be
V_{us}\approx\frac{\delta}{\epsilon}~~~~~~~~~ V_{ub}\approx \delta~~~~~~~~~~ V_{cb}\approx \epsilon~~~.
\ee
We can match almost all experimental data, by taking
$\epsilon$ of order $\lambda^2$ and $\delta$ of order
$\lambda^3\div\lambda^4$. We observe that the choice 
$\delta=\lambda^3$ seems more appropriate to reproduce
the observed quark mixing angles, needed for our estimate
of the proton lifetime.
For neutrinos, in both options I and II, we should invoke some
additional accidental relations \cite{review}, not automatically guaranteed
by the textures in eq. (\ref{denu}). In case I, known as
``semianarchy'', the determinant 
of the 23 block in $m_\nu$, which is generically of order one, 
should be tuned around $m_2/m_3\approx\sqrt{\Delta m^2_{sol}/
\Delta m^2_{atm}}\approx 0.1\div 0.2$. We should also enhance
somewhat the solar mixing angle.
In case II, denoted as ``anarchy'' \cite{anarchy}, 
we need again the same cancellation in the 23 determinant
and we should adequately suppress the $\theta_{13}$ angle.
The overall picture is not entirely satisfactory, though, at least 
from a technical view point, the modest required tuning can be easily 
realized by exploiting the freedom in the dimensionless coefficients 
that multiply each matrix element in the above textures. 
 
When computing the amplitudes for proton decay, we will need
the transformations mapping fermions from the interaction basis 
into the mass eigenstate basis. We can express these transformations
as:
\be
\begin{array}{cccc}
u\to L_u u & d\to L_d d & e\to L_e e & \nu \to L_\nu \nu\\
u^c\to R_u^\dagger u^c & d^c\to R_d^\dagger d^c &
e^c\to R_e^\dagger e^c &
\end{array}~~~~~,
\label{map}
\ee
where in the left-hand (right-hand) side fields are in the interaction 
(mass eigenstate) basis
and $L$, $R$ are 3$\times$ 3 unitary matrices. The quark
mixing matrix $V_{CKM}$ and the lepton one $U_{PMNS}$
are given by:
\be
V_{CKM}=L_u^\dagger L_d~~~,~~~~~~~U_{PMNS}=L_e^\dagger L_\nu~~~.
\ee
The unitary matrices $L$ and $R$ can be estimated from eqs.
(\ref{newmu},\ref{newdenu}) and the action in eq. (\ref{yuk}).
We find:
\be
L_u\equiv R_u\approx L_d\approx R_e\approx
\left(
\begin{array}{ccc}
1 & \lambda & \lambda^3 \\
\lambda & 1 & \lambda^2 \\
\lambda^3 & \lambda^2 & 1
\end{array}
\right)
\div
\left(
\begin{array}{ccc}
1 & \lambda^2 & \lambda^4 \\
\lambda^2 & 1 & \lambda^2 \\
\lambda^4 & \lambda^2 & 1
\end{array}
\right)~~~,
\label{luld}
\ee
letting $\delta$ vary in the range $\lambda^3\div \lambda^4$.
When $\delta=\lambda^4$ some accidental enhancement is required to
correctly reproduce $V_{us}$ and $V_{ub}$. We have also
\be
\begin{array}{c}
{\tt option~ I}\\
\\
L_e\approx R_d
\approx\left(
\begin{array}{ccc}
1 & \lambda^2 & \lambda^2\\
\lambda^2 & 1 & 1\\
\lambda^2 & 1 & 1
\end{array}
\right)\\
\\
L_\nu
\approx\left(
\begin{array}{ccc}
1 & \lambda & \lambda^2\\
\lambda & 1 & 1\\
\lambda^2 & 1 & 1
\end{array}
\right)\\
\end{array}~~~~~~~~~~~~~~~~~~~~~~~~~
\begin{array}{c}
{\tt option~ II}\\
\\
L_e\approx R_d\approx L_\nu
\approx\left(
\begin{array}{ccc}
1 & 1 & 1\\
1 & 1 & 1\\
1 & 1 & 1
\end{array}
\right)\\
\\
\end{array}~~~~~,
\label{lelnu}
\ee
\\[0.3cm]
where, in option I, we have assumed that the determinant of the 
23 block of $w$ is of order $\lambda$.
Moreover, in option II, some accidental cancellation should guarantee
$\theta_{13}\approx\lambda$.
Eqs. (\ref{luld},\ref{lelnu}) are basic inputs in our estimates
of the proton lifetime. 

\subsection{Supersymmetry breaking}

In the model under discussion, baryon violating processes are dominated by dimension six operators 
originating from vector boson exchange and, in first approximation,
the proton lifetime is not sensitive to the breaking of the
residual four-dimensional N=1 SUSY. Nevertheless
a dependence on the light supersymmetric spectrum arises from
the threshold corrections to gauge coupling unification, around
the SUSY breaking scale. As we shall see in detail in section 4,
the compactification scale that controls proton decay amplitudes
is determined by gauge coupling unification and, in a phenomenological
analysis, is slightly affected by the light supersymmetric particles.
\begin{table}[h!]
%\caption{\small $Spettri$ $Susy$}
\begin{center}
\label{SpettriSusy}
\vspace{0.1cm}
\begin{tabular}{|c|c|c|c|c|c|c|c|c|c|c|}   

\hline
& & & & & & & & & &  \\
{\tt particle}& SPS1a &SPS1b & SPS2 & SPS3 & SPS4 & SPS5 & SPS6 & SPS7 & SPS8 & SPS9 \\

\hline  
& & & & & & & & & &  \\
$\tilde{g}$ & $600$ & $925$ & $780$ & $910$ & $720$ & $710$ & $710$ & $925$ & $825$ & $1280$\\
\hline  

\hline  
& & & & & & & & & &  \\
$\tilde{w}$ & $200$ & $320$ & $240$ & $320$ & $240$ & $240$ & $240$ & $260$ & $250$ & $550$ \\
\hline  

\hline  
& & & & & & & & & &  \\
$\tilde{h}$ & $360$ & $505$ & $300$ & $530$ & $400$ & $640$ & $425$ & $385$ & $430$ & $1050$\\
\hline  

\hline  
& & & & & & & & & &  \\
{\tt extra H} & $400$ & $543$ & $1450$ & $575$ & $410$ & $690$ & $470$ & $385$ & $520$ & $1080$ \\
\hline  

\hline  
& & & & & & & & & & \\
$\tilde{q}$ & $525$ & $835$ & $1500$ & $820$ & $730$ & $650$ & $650$ & $845$ & $1080$ & $1245$\\
\hline  

\hline  
& & & & & & & & & &  \\
$\tilde{u^c}$ & $525$ & $800$ & $1500$ & $800$ & $715$ & $620$ & $620$ & $830$ & $1030$ & $1245$ \\
\hline  

\hline  
& & & & & & & & & &  \\
$\tilde{d^c}$ & $547$ & $800$ & $1500$ & $800$ & $715$ & $620$ & $620$ & $830$ & $1030$ & $1245$ \\
\hline  

\hline  
& & & & & & & & & &  \\
$\tilde{l}$ & $200$ & $345$ & $1450$ & $800$ & $450$ & $260$ & $260$ & $255$ & $350$ & $390$ \\
\hline 

\hline  
& & & & & & & & & & \\
$\tilde{e^c}$ & $140$ & $250$ & $1450$ & $280$ & $415$ & $190$ & $230$ & $125$ & $175$ & $360$ \\
\hline  

\end{tabular}
\caption{\small Spectrum of SUSY particles. Masses are in GeV. The values are taken from ref. 
\cite{SPS} and refer to different realistic scenarios. Here 
they are given in an SU(2) invariant approximation.}
\end{center}
\end{table}

There are several SUSY breaking mechanisms that can be adapted 
to the present setup. It is possible to break SUSY by non-trivial
boundary conditions on the bulk superfields \cite{SSSB,SSSB1}.
We will denote such possibility as Scherk-Schwarz SUSY breaking (SSSB).
In this case, at the tree-level,
the splitting within N=1 supermultiplets only affects bulk fields and is proportional to $1/R$,
which in the present context is close to the unification scale.
Therefore, to keep superparticles at the TeV scale, it is necessary
to have a tiny dimensionless SUSY breaking parameter \cite{bhn}.
SUSY can also be broken by an intrinsic four-dimensional
mechanism on either of the two branes. The SUSY breaking sector and 
the transmission of SUSY breaking to the
observable sector are notoriously a source of ambiguities
and phenomenological problems. Here we are only interested 
in the estimate of the theoretical uncertainty on proton lifetime coming
from our ignorance about the SUSY breaking mechanism.
We think that, from this viewpoint, the best way of parametrizing 
such an ignorance is to assume a variety of supersymmetric particle spectra 
and, for each of them, evaluate the proton decay rate for the various channels.
The comparison among the results corresponding to different spectra
will quantify the theoretical uncertainty we are interested in.
As we will see, this uncertainty is of a certain relevance but it
is not the dominant one in our problem.

In our numerical estimates we have adopted the ten spectra listed
in table 4. These spectra have been adapted from ref. \cite{SPS}
where they correspond to the so-called Snowmass Points and Slopes (SPS),
a set of benchmark points and parameter lines in the MSSM parameter space
corresponding to different scenarios. We have simplified the spectra
of ref. \cite{SPS}, by taking an SU(2) limit where all left-handed sfermions
are degenerate. Similarly winos and higgsinos are taken as unmixed
mass eigenstates. In addition, in our option II for matter fields, we
have also considered the spectrum of ref. \cite{hn3}, obtained by
breaking SUSY through boundary conditions for the bulk fields.
This spectrum is listed in table 5.

\begin{table}
\begin{center}
\label{SpettriSusy2}
\vspace{0.1cm}
\begin{tabular}{|c|c|c|c|c|c|c|c|c|}   

\hline
& & & & & & & & \\
$\tilde{g} $ & $\tilde{w}$ & $\tilde{h}$ & ${\tt extra H}$ &$\tilde{q}$ & $\tilde{u^c}$ & $\tilde{d^c}$ & 
$\tilde{l}$ & $\tilde{e^c}$\\
& & & & & & & & \\
\hline

\hline
& & & & & & & & \\
& & & & $701$ & $675$ & $602$ & $209$ & $317$\\
$699$ & $251$ & $427$ & $555$ & & & & & \\
& & & & $610$ & $425$ & $563$ & $214$ & $106$\\
& & & & & & & & \\
\hline
\end{tabular}   
\caption{\small Masses in GeV of SUSY particles within SSSB, 
from ref. \cite{hn3}, in a $SU(2)$ invariant approximation.
In the sfermion sector the first row refers to first and second generation, while the second row
corresponds to the third generation.}
\end{center}
\end{table}

%%%%%%%%%%%%%%%%%%%%% 3. DIMENSION 6 OPERATORS %%%%%%%%%%%%%%%%%%%%%%%%%%%%

\section{Dimension 6 operators for proton decay}

The interactions contributing to proton decay are those between the
gauge vector bosons $X\equiv A^{\hat{a}}_\mu$ and the N=1 chiral multiplets
on the brane at $y=0$.
In option 0, where all matter fields are at $y=0$,
the effective lagrangian for proton decay is formally identical
to that from gauge boson exchange in minimal SUSY SU(5):
\bea
{\cal L}_p&=& -\frac{g_U^2}{2 M_X^2} \left[
\left(\delta_{ij}\delta_{kl}+(V_{CKM})_{il}(V_{CKM}^\dagger)_{kj}\right)~
\overline{u^c_i} \gamma^\mu u_j \cdot
\overline{e^c_k} \gamma_\mu d_l\right.\nn\\
&-&\left.
\overline{u^c_i} \gamma^\mu u_i \cdot
\overline{d^c_j} \gamma_\mu e_j
+
\overline{u^c_i} \gamma^\mu (V_{CKM})_{ij} d_j \cdot
\overline{d^c_k} \gamma_\mu \nu_k
\right]~~~,
\label{lagpd0}
\eea
where all fermions are mass eigenstates (neutrinos are taken as massless) 
and $i,j,k=1,2,3$ are generation indices. Colour indices are understood. 
The dimensionless quantity 
\be
g_U\equiv g_5 \frac{1}{\sqrt{2 \pi R}}
\ee
is the four-dimensional gauge coupling of the gauge vector bosons
zero modes. The combination
\be
M_X=\frac{2 M_c}{\pi}~~~,
\ee
proportional to the compactification scale
\be
M_c\equiv \frac{1}{R}~~~,
\ee
is an effective gauge vector boson mass arising from the sum over all the Kaluza-Klein levels:
\be
\sum_{n=0}^\infty\frac{1}{(2 n+1)^2 M_c^2}=\frac{1}{2 M_X^2}~~~.
\ee
When expressing interaction eigenstates in terms of mass eigenstates,
through eq. (\ref{map}), we have exploited the minimal SU(5)
relations $R_u^\dagger=L_u$, $R_d^\dagger=L_e$, $R_e^\dagger=L_d$,
which hold also in option I.

When considering option I and II the interactions between $A^{\hat{a}}_\mu$ 
and the matter multiplets on the brane at $y=0$ are
\be
{\cal S}_{\Delta B \neq 0} = - \frac{g_5}{\sqrt{2}} \int d^4x
{\cal A}^{(+-)\alpha I}_{\mu}(x,0) J_{\alpha I}^{\mu}(x)+ h.c.~~~, 
\ee
where ${\cal A}^{(+-)\alpha I}_{\mu}(x,y)$ is the five dimensional
gauge vector boson with quantum numbers $({\bar 3},2,5/6)$ under 
SU(3)$\times$ SU(2) $\times$ U(1), $\alpha$ and $I$ are
color and SU(2) indices respectively.
The currents $J_{\alpha I}^{\mu}$ are given by:
\bea
J_{\alpha I}^{\mu}&=&
\epsilon_{\alpha\beta\gamma}\,\epsilon_{IJ} 
\overline{q}^{\beta J}_3\,\gamma^\mu\, u^{c \gamma}_3 + 
\overline{e^c}_3\, \gamma^\mu\, q_{\alpha I 3} - 
\overline{d^c}_{\alpha a}\, \gamma^\mu\, l_{I a}\nn\\
&=&
\overline{q}_3\,\gamma^\mu\, u^c_3 + 
\overline{e^c}_3\, \gamma^\mu\, q_3 - 
\overline{d^c}_{a}\, \gamma^\mu\, l_{a}~~~,
\eea
where in the second line gauge indices have been abolished
and only generation labels are presents: the sum over $a$ 
is restricted to the multiplets $F_a$
living on the brane at $y=0$. All fermions $q$, $l$, $u^c$, $d^c$ and $e^c$ are left-handed 
in our notation.
If we insert the expansion of eq. (\ref{phi+-exp}) and we integrate out the
super heavy gauge bosons in the limit of vanishing momenta for the
light particles, we obtain the four-fermion Lagrangian:
\be
{\cal L}_p= -\frac{g_U^2}{2 M_X^2} \left(
\overline{u^c_3} \gamma^\mu q_3 \cdot
\overline{e^c_3} \gamma_\mu q_3
-
\overline{u^c_3} \gamma^\mu q_3 \cdot
\overline{d^c_a} \gamma_\mu l_a
\right)~~~.
\label{lagk}
\ee
By expressing it in terms of
mass eigenstates through the transformations of eq. (\ref{map}),
we get:
\be
{\cal L}_p= -\frac{g_U^2}{2 M_X^2} \left(
C^1_{ij}~
\overline{u^c} \gamma^\mu u \cdot
\overline{e^c_i} \gamma_\mu d_j
-
C^2_{ij}~
\overline{u^c} \gamma^\mu u \cdot
\overline{d^c_j} \gamma_\mu e_i
+
C^3_{ijk}~
\overline{u^c} \gamma^\mu d_i \cdot
\overline{d^c_j} \gamma_\mu \nu_k
\right)~~~.
\label{lagpd}
\ee
Here $u$ denotes the lightest up type quark, $i,j=1,2$ and $k=1,2,3$.
The coefficients $C$'s are given by:
\bea
C^1_{ij}&=&2 (R_u)_{13} (L_u)_{31} (R_e)_{i3} (L_d)_{3j}\nn\\
C^2_{ij}&=& (R_u)_{13} (L_u)_{31} (R_d)_{ja} (L_e)_{ai}\nn\\
C^3_{ijk}&=& (R_u)_{13} (L_d)_{3i} (R_d)_{ja} (L_\nu)_{ak}
\label{coefc}
\eea
where, as above, the sum over $a$ is restricted to the multiplets $F_a$
living on the brane at $y=0$. From the previous equations
we can easily estimate the leading contribution to the decay amplitudes.
In option 0 the result coincides with those due to $X$ boson
exchange in minimal SUSY SU(5), with the substitution $M_X\to
2 M_c/\pi$. In particular the dominant decay mode
is $\pi^0 e^+$, whose branching ratio is approximately $30\div 40\%$
\cite{cfz}.

In option I and II, 
by neglecting any numerical coefficients and any subleading contributions,
we obtain the results listed in table 6. From table 6 we see that the contribution from the first term in
the expression of the Lagrangian (\ref{lagk}), proportional
to the coefficients $C^1_{ij}$, is always negligible. This is due to the
high suppression brought by the mixing matrices when converting
third generation fields into light fermions.
From this point of view the suppression is less severe for the second
term in eq. (\ref{lagk}), which gives origin to $C^2_{ij}$ and 
$C^3_{ijk}$. When $F_1$ is taken as a bulk field (option I),
the dominant operators are
$uus\mu$ and $uds\nu_{\mu,\tau}$, with a flavour suppression of order
$\lambda^{6\div 8}$. This favors the decay channels $p\to K^0 \mu^+$
and $p\to K^+ {\bar\nu}$.
When $F_1$ is chosen as a brane field (option II),
the dominant operator is $uds\nu$, with a neutrino of whatever type
and a suppression of order $\lambda^{5\div 6}$.
The preferred decay channel is $K^+ {\bar\nu}$.
All the other possible
operators contributing to proton decay, 
listed in table 6, are depleted
with respect to the dominant one, with a parametric suppression
$\lambda^{6\div 8}$.
\begin{table}
{\begin{center}
\begin{tabular}{|c|c|c|}   
\hline
& & \\                         
{\tt operator} & {\tt option I} & {\tt option II} \\ 
& &\\
\hline
& & \\                         
${\tt u u d e}$ & $C^2_{11}\approx \lambda^{10\div 12}$ & 
$C^2_{11}\approx \lambda^{6\div 8}$\\ 
& &\\
\hline
& & \\                         
${\tt u u d \mu}$ & $C^2_{21}\approx \lambda^{8\div 10}$ & 
$C^2_{21}\approx \lambda^{6\div 8}$\\ 
& &\\
\hline
& & \\                         
${\tt u u s e}$ & $C^2_{12}\approx \lambda^{8\div 10}$ & 
$C^2_{12}\approx \lambda^{6\div 8}$\\ 
& &\\
\hline
& &\\
${\tt u u s \mu}$ & $C^2_{22}\approx \lambda^{6\div 8}$ & 
$C^2_{22}\approx \lambda^{6\div 8}$\\ 
& &\\
\hline
\hline
& &\\
$\begin{array}{c}
{\tt u d d \nu_e}\\
\\
{\tt u d d \nu_{\mu,\tau}}\\
\\
\end{array}$
&
$\begin{array}{c}
C^3_{111}\approx\lambda^{9\div 11}\\
\\
C^3_{11k}\approx\lambda^{8\div 10}\\
{\scriptstyle (k=2,3)} 
\end{array}$
&
$\begin{array}{c}
C^3_{11k} 
\approx\lambda^{6\div 8}\\
{\scriptstyle (k=1,2,3)}
\end{array}$\\
& &\\
\hline
& &\\
$\begin{array}{c}
{\tt u d s \nu_e}\\
\\
{\tt u d s \nu_{\mu,\tau}}\\
\\
\end{array}$
&
$\begin{array}{c}
C^3_{121}\approx\lambda^{7\div 9}\\
\\
C^3_{12k}\approx\lambda^{6\div 8}\\
{\scriptstyle (k=2,3)} 
\end{array}$
&
$\begin{array}{c}
C^3_{21k} 
\approx\lambda^{5\div 6}\\
{\scriptstyle (k=1,2,3)}
\end{array}$\\
& &\\
\hline
\end{tabular} 
\end{center}}
\caption{Leading contributions to the different decay amplitudes
from the Lagrangian of eq. (\ref{lagpd}). Only the order of magnitude 
in $\lambda$ is displayed. Powers of $\lambda$ vary in the range displayed
in the table depending on the choice $\delta\approx \lambda^3\div\lambda^4$.}
\end{table}
This discussion shows that, even in the most favorable case, we should expect
a large flavor suppression in the decay amplitude, with respect to 
option 0 where essentially
no small angles are required to proceed from flavour to the relevant 
mass eigenstates. Numerically $\lambda^{6\div 5}\approx 
(1\div 5)\times 10^{-4}$
and it would seem quite difficult to rescue such a small numerical factor
and end up with a measurable effect.
In the next section we will carefully analyze gauge coupling unification
in this class of models, in order to quantify the overall coefficient 
$g_U^2/(2 M_X^2)$ and its impact on the proton lifetime. 

%%%%%%%%%%%%%%%%% 4. GAUGE COUPLING UNIFICATION %%%%%%%%%%%%%%%%%%%%%%%%

\section{Gauge coupling unification}

The overall strength of proton decay amplitudes is governed by
the masses of the gauge vector bosons $A^{\hat{a}}_\mu$, which, in the
present model is proportional to the compactification scale $M_c$.
The latter can be estimated by analyzing the unification of gauge couplings.
The low-energy coupling constants $\alpha_i(m_Z)$ $(i=1,2,3)$
in the $\overline{MS}$ scheme are related to the unification scale 
$\Lambda_U$, the common value 
$\alpha_U=g_U^2/(4\pi)$ at $\Lambda_U$ and the compactification scale $M_c$
by the renormalization group equations:
\be
\frac{1}{\alpha_i(m_Z)}=\frac{1}{\alpha_U}
+\frac{b_i}{2\pi}\log \left(\frac{\Lambda_U}{m_Z}\right)
+ \delta_i~~~.
\label{rge}
\ee
Here $b_i$ are the coefficient of the SUSY $\beta$ functions in
the one-loop approximation:
\be
\left(
\begin{array}{c}
b_1\\b_2\\b_3
\end{array}
\right)=  
\left(
\begin{array}{c}
33/5\\1\\-3
\end{array}
\right)
~~~,
\ee
for 3 generations and 2 light Higgs SU(2) doublets. We recall that
$g_1$ is related to the hypercharge coupling constant $g_Y$ by
$g_1=\sqrt{5/3}~ g_Y$. In eq. (\ref{rge}), $\delta_i$ stand for non-leading
contributions and depend upon $M_c$. More precisely:
\be
\delta_i=\delta^{(2)}_i+\delta^{(l)}_i+\delta^{(h)}_i+\delta^{(b)}_i~~~,
\label{deltas}
\ee 
where $\delta^{(2)}_i$ represent two-loop running effects,
$\delta^{(l)}_i$ are light threshold corrections at the SUSY breaking
scale, $\delta^{(h)}_i$ are heavy threshold corrections at the compactification
scale $M_c$ and $\delta^{(b)}_i$ denote SU(5) breaking contributions
from the brane at $y=\pi R/2$. By setting $\delta_i=0$ we go back to the leading
order approximation where the three gauge coupling constants exactly
unify at the scale $\Lambda_U$. In such an approximation, 
from the experimental values of $\alpha_{em}(m_Z)$ and $\sin^2\theta_W(m_Z)$
we can derive $\Lambda_U$, $\alpha_U$ and $\alpha_3(m_Z)$.
In the present context, we will use equations (\ref{rge}) to evaluate
$M_c$ contained in $\delta^{(h)}_i$, starting from the experimental values
$\alpha_{em}(m_Z)$, $\sin^2\theta_W(m_Z)$ and $\alpha_3(m_Z)$
and from estimates of $\delta^{(2)}_i$, $\delta^{(l)}_i$ and $\delta^{(b)}_i$.
We use the experimental inputs \cite{pdg}:
\bea
\alpha_{em}^{-1}(m_Z)&=&127.906\pm 0.019\nn\\
\sin^2\theta_W(m_Z)&=&0.2312\pm 0.0002\nn\\
\alpha_3(m_Z)&=&0.1187\pm 0.0020~~~~~.
\label{inputrge}
\eea

\subsection{Two-loop}

We include the two-loop corrections coming from the gauge sector 
\cite{twoloop}:
\be
\delta_i^{(2)}=\frac{1}{\pi}\sum_{j=1}^3
\frac{b_{ij}}{b_j}
\log\left[1+b_j\left(\frac{3-8\sin^2\theta_W}{36\sin^2\theta_W-3}\right)\right]~~~,
\ee
where the matrix $b_{ij}$ is given by:
\be
b_{ij}=
\left(
\begin{array}{ccc}
199/100& 27/20& 22/5\\
9/20& 25/4& 6\\
11/20& 9/4& 7/2
\end{array}
\right)~~~.
\ee

\subsection{Light thresholds}

The threshold effects from the SUSY breaking scale are provided by 
\cite{susyth}:
\be
\delta_i^{(l)}=-\frac{1}{\pi}\sum_j b_i^{(l)}(j) 
\log\left(\frac{m_j}{m_Z}\right)~~~
\label{light}
\ee
where the index $j$ runs over the spectrum of SUSY particles of masses $m_j$
and extra Higgses and the coefficients $b_i^{(l)}(j)$ are given in table 7.
We work in the approximation where SU(2) breaking effects are neglected.
The meaning of $b_i^{(l)}(j)$ is very simple and can most easily be 
captured in the approximation where all particles have a common mass
$m_{SUSY}$. In such a case, from table 7 and eq. (\ref{light}) we obtain:
\be
\delta_1^{(l)}=-\frac{5}{4\pi}\log\frac{m_{SUSY}}{m_Z}~~~
\delta_2^{(l)}=-\frac{25}{12\pi}\log\frac{m_{SUSY}}{m_Z}~~~
\delta_3^{(l)}=-\frac{2}{\pi}\log\frac{m_{SUSY}}{m_Z}~~~
\ee
\\[0.1cm]
{\begin{center}
\begin{tabular}{|c|c|c|c|}   
\hline
& & &\\  
{\tt particle}& $b_1^{(l)}(j)$& $b_2^{(l)}(j)$& $b_3^{(l)}(j)$\\
& & &\\
\hline
& & &\\
$\tilde{g}$& 0& 0& 1\\  
& & &\\
\hline
& & &\\
$\tilde{w}$& 0& 2/3& 0\\  
& & &\\
\hline
& & &\\
$\tilde{h}$& 1/5& 1/3& 0\\  
& & &\\
\hline
& & &\\
$\tilde{q}$& 1/60& 1/4& 1/6\\  
& & &\\
\hline
& & &\\
$\tilde{u^c}$& 2/15& 0& 1/12\\  
& & &\\
\hline
& & &\\
$\tilde{d^c}$& 1/30& 0& 1/12\\  
& & &\\
\hline
& & &\\
$\tilde{l}$& 1/20& 1/12& 0\\  
& & &\\
\hline
& & &\\
$\tilde{e^c}$& 1/20& 1/12& 0\\  
& & &\\
\hline
& & &\\
{\tt extra H}&1/20& 1/12& 0\\
& & &\\
\hline
\end{tabular}
\end{center}}
\vspace{3mm}
Table 7. Coefficients $b_i^{(l)}(j)$ for SUSY particles and extra Higgses.
All SU(2) breaking effects are neglected. Sfermion contributions in the table 
refer to one generation.
\\[0.3cm]
If we add these contributions to the leading order terms, we get
\be
\frac{1}{\alpha_i(m_Z)}=\frac{1}{\alpha_U}
+\frac{b^{SM}_i}{2\pi}\log \left(\frac{m_{SUSY}}{m_Z}\right)
+\frac{b_i}{2\pi}\log \left(\frac{\Lambda_U}{m_{SUSY}}\right)~~~,
\ee
where $(b_1^{SM},b_2^{SM},b_3^{SM})=(41/10,-19/6,-7)$ are the coefficients
of the $\beta$ functions in the one-loop approximation in the standard model.
In this simplified case the running of the gauge coupling constants
is standard from $m_Z$ to $m_{SUSY}$ and supersymmetric from $m_{SUSY}$ up to
the unification scale $\Lambda_U$.
As explained in section 2.5, we do not adopt a specific scheme
for SUSY breaking. In the model under discussion, where 
dimension 5 baryon-violating operators are forbidden, the coefficients 
$\delta_i^{(l)}$ are the only quantities entering the evaluation of 
the proton lifetime that are sensitive to the details of SUSY breaking.
To the purpose of providing a realistic estimate
of the proton lifetime, we can better size the uncertainty from the 
SUSY breaking sector by considering the variety of sparticle spectra
listed in tables 4 and 5, coming from several mechanisms.  
As a consequence, the contribution $\delta_i^{(l)}$ varies in a range 
that describes our ignorance about the SUSY breaking sector.

\subsection{Brane kinetic terms}

The contributions $\delta_i^{(b)}$ in eq. (\ref{deltas}) come from
kinetic energy terms for the gauge bosons of SU(3), SU(2) and U(1) 
on the brane at $y=\pi R/2$. These terms, which break SU(5), are allowed
by the symmetries of the theory and, even if we set them to zero at the 
tree-level, they are generated by radiative corrections \cite{cprt,ggh}.
The net effect of these terms is to modify the boundary value of the 
gauge coupling constants $g_i(\Lambda)$ at the cut-off scale $\Lambda$:
\be
\frac{1}{g_i^2(\Lambda)}=\frac{2\pi R}{g_5^2(\Lambda)}+
\frac{1}{g_{(b)i}^2}
\label{glambda}
\ee
where $g_5(\Lambda)$ is the SU(5)-invariant gauge coupling constant, 
having mass dimension -1/2, coming from the bulk kinetic terms of 
the gauge bosons, while $g_{(b)i}$ are dimensionless gauge couplings
arising from independent kinetic terms of the SM gauge fields
on the SU(5) breaking brane at $y=\pi R/2$.
If the SU(5) breaking terms $1/g_{(b)i}^2$ where similar in size to the
symmetric one, we would loose any predictability and we could not hope
to reasonably constrain the compactification scale $M_c$, through
the evolution of gauge couplings. It has been observed that a predictive
framework is recovered if we assume that the theory is strongly coupled
at the cut-off scale $\Lambda_U$ \cite{n2}. In such a case, dimensional analysis
suggests that $g_5^2(\Lambda)\approx 16 \pi^3/\Lambda$, 
$g_{(b)i}^2\approx 16\pi^2$ and eq. (\ref{glambda}) becomes:
\be
\frac{1}{g_i^2(\Lambda)}\approx\frac{\Lambda R}{8 \pi^2}+
O(\frac{1}{16\pi^2})~~~.
\label{Bandb}
\ee
Since $\Lambda R\gg1$ is expected for the consistency of the model
and $\Lambda R\approx O(100)$ is welcome from the view point of
fermion masses, we see that the SU(5) invariant component indeed dominates
over the non-symmetric one and that the theory predicts gauge couplings
of order one at the cut-off scale. In what follows we will adopt
the assumption of strong gauge coupling at $\Lambda$.
Of course, since our quantitative analysis will allow to explicitly
evaluate both $R$ and $\Lambda$, we will be able to check
the validity of this assumption.
For our numerical analysis, we will regard $\delta_i^{(b)}$ as
random numbers with a flat distribution in the interval $(-1/2\pi,+1/2\pi)$:
\be
\delta_i^{(b)}\in \left[-\frac{1}{2\pi},+\frac{1}{2\pi}\right]~~~.
\label{random}
\ee
Given our ignorance about the ultraviolet completion of our model,
where we could predict the parameters $\delta_i^{(b)}$,
each value in the above interval has the same ``probability''.
This will induce an uncertainty in the determination of the
compactification scale $M_c$, which, as we shall see, represents 
the largest theoretical error. 

\subsection{Heavy Thresholds}

Two different approaches to the evaluation of the heavy threshold effects
exist in the literature. In ref. \cite{hn1}, which we will refer to 
as {\em HN},
these effects are evaluated
in a leading logarithmic approximation for the particles whose masses
are smaller than the cut-off scale $\Lambda$. These states
belong to the Kaluza-Klein towers of the gauge bosons 
and the matter hypermultiplets. The result is:
\be
\delta^{(h)HN}_i=
\frac{\gamma_i}{2\pi}
\sum_{n=0}^N \log\frac{\Lambda}{(2n+2)M_c}+
\frac{\eta_i}{2\pi}
\sum_{n=0}^N \log\frac{\Lambda}{(2n+1)M_c}
\label{hn1}
\ee
where the coefficients $\gamma_i$ and $\eta_i$ are listed in table 8. 

The sums stop at $N$ such that
$(2N +2) M_c$ is the Kaluza-Klein level closest to, but still
smaller than, the cut-off $\Lambda$:
\be
(2 N+2)\approx\frac{\Lambda}{M_c}~~~.
\ee
It is convenient to recast
eq. (\ref{hn1}) in the form
\be
\delta^{(h)HN}_i=\delta^{(h)}_U+
\frac{\sigma_i}{2\pi}
\sum_{n=0}^N \log\frac{(2n+1)}{(2n+2)}~~~,
\label{hn2}
\ee
where $\sigma_i$ are given in table 8
and $\delta^{(h)}_U$ is a universal contribution that
can be absorbed into $\alpha_U$:
\be
\delta^{(h)}_U= 
\frac{1}{2\pi}
(-4+n_{\bar{5}}+3 n_{10})\sum_{n=0}^N \log\frac{\Lambda^2}{(2n+1)(2n+2)M_c^2}~~~,
\ee
where $n_{\bar{5}}$ and $n_{10}$ are the number of matter hypermultiplets 
transforming as $\bar{5}$ and $10$ of SU(5).   
For large $N$, that is for $\Lambda R\gg 1$
\be
\sum_{n=0}^N \log\frac{(2n+1)}{(2n+2)}
\approx
-\frac{1}{2}\log(N+1)-\frac{1}{2}\log\pi
\approx
-\frac{1}{2}\log\frac{\Lambda}{M_c}-\frac{1}{2}\log\frac{\pi}{2}~~~.
\ee
In this limit:
\be
\delta^{(h)HN}_i=
-\frac{\sigma_i}{4\pi}\log\frac{\Lambda}{M_c}+...
~~~,
\label{hn3}
\ee
where dots stand for universal contributions. From eq. (\ref{hn1},\ref{hn2})
we also see that, up to an irrelevant universal contribution redefining
the initial condition $\alpha_U$, all the effect comes from the shift
between even and odd Kaluza-Klein levels that removes the degeneracy within
full SU(5) multiplets.

In ref. \cite{cprt}, referred to as {\em CPRT}, $\delta^{(h)}_i$ are estimated within an effective
lagrangian approach. The full towers of Kaluza-Klein states
are integrated out in the higher-dimensional model and an effective 
theory at the scale $M_c$ is obtained. The gauge couplings at the
$M_c$ are related to the gauge couplings at the cut-off scale
by the threshold corrections: 
\be
\delta^{(h)CPRT}_i=
-\frac{\gamma_i}{4\pi}\log\frac{\Lambda}{M_c}
-\frac{\gamma_i}{4\pi}(\log\pi+1-{\cal I})
-\frac{\eta_i}{4\pi}\log 2
+\frac{\delta^{c}_i}{2\pi}
\label{cprt}
\ee
where
\be
{\cal I}=\frac{1}{2}\int_1^{+\infty} dt (t^{-1}+t^{-1/2})
(\sum_{n=-\infty}^{+\infty}e^{-\pi t n^2}-1)\approx 0.02~~~,
\ee
and $\delta^{c}_i$ are conversion factors from the
$\overline{\rm DR}$ to the $\overline{\rm MS}$
scheme \cite{conv} also given in table 8.

In the limit $\Lambda\gg M_c$, the term proportional
to $\log(\Lambda/M_c)$ dominates both $\delta^{(h)HM}_i$
in eq. (\ref{hn3}) and $\delta^{(h)CPRT}_i$ in eq. (\ref{cprt}).
Actually, since $\gamma_i=\sigma_i-4$, up to an irrelevant
universal contribution, the two expressions of eqs. (\ref{hn3})
and (\ref{cprt}) coincide in this limit.
In our analysis we have used both $\delta^{(h)HM}_i$ and $\delta^{(h)CPRT}_i$
to estimate the compactification scale $M_c$. The difference between
the values of $M_c$ obtained in these two approaches is part
of the theoretical error affecting our estimates. 
\\[0.1cm]
{\begin{center}
\begin{tabular}{|c||c|c|c|c|}   
\hline
& & & & \\
$i$& $\gamma_i$& $\eta_i$& $\sigma_i$& $\delta^c_i$\\
& & & &\\
\hline
\hline
& & & &\\
1& 6/5& -46/5& 26/5&  0\\
& & & &\\
\hline
& & & &\\
2& -2& -6& 2& 1/3\\
& & & &\\
\hline
& & & &\\
3& -6& -2& -2& 1/2\\
& & & &\\
\hline
\end{tabular}
\end{center}}
\vspace{3mm}
Table 8. Coefficients for the heavy threshold corrections due to the
gauge and Higgs sectors
eqs. (\ref{hn1},\ref{hn2},\ref{cprt}), from ref. \cite{hn1} and \cite{cprt}.
If $n_{\bar{5}}$ and $n_{10}$ matter hypermultiplets transforming as
$\bar{5}$ and $10$ of SU(5) are present, the additional contribution
$(n_{\bar{5}}+3 n_{10})$ should be added to both $\gamma_i$ and $\eta_i$,
for all $i=1,2,3$. (Two copies of $n$ SU(5) multiplets give rise to 
$n$ zero modes in the present model.)
\\[0.3cm]
We can have a qualitative understanding of the non-leading effects
by looking at the strong coupling constant $\alpha_3(m_Z)$.
At leading order, defined by setting $\delta_i$ to zero in eq.   
(\ref{rge}), from the input values of $\alpha_{em}(m_Z)$ and
$\sin^2\theta_W(m_Z)$ we obtain the prediction:
\be
\alpha_3^{LO}(m_Z)\approx 0.118~~~,
\label{lo}
\ee
in excellent agreement with the experimental value.
When we turn on the non-leading corrections $\delta_i$, the predicted
value of $\alpha_3(m_Z)$ is changed into:
\bea
\alpha_3(m_Z)&=&\alpha_3^{LO}(m_Z)\left[1-\alpha_3^{LO}(m_Z)\delta_s\right]\nn\\
\delta_s&=&\frac{1}{7}\left(5\delta_1-12\delta_2+7\delta_3\right)~~~,
\eea
where the combination $\delta_s$ is the sum
of $\delta_s^{(2)}$, the two-loop correction, $\delta_s^{(l)}$, the
light threshold correction, $\delta_s^{(b)}$, the brane contribution and $\delta_s^{(h)}$, 
the heavy threshold correction. From our previous discussion, these contributions
are readily evaluated and we obtain:
\bea
\delta_s^{(2)}&\approx& -0.82\nn\\
\delta_s^{(l)}&\approx& -0.50+\frac{19}{28\pi}\log\frac{m_{SUSY}}{m_Z}\nn\\
\delta_s^{(b)}&\approx& \pm \frac{1}{2\pi}\nn\\
\delta_s^{(h)}&\approx& \frac{3}{7\pi}\log\frac{\Lambda}{M_c}~~~,
\eea
where we have considered a `typical' spectrum of supersymmetric particles 
and only the leading logarithmic correction from heavy thresholds.
The combined effect of two-loop corrections and light thresholds would
raise the prediction of $\alpha_3(m_Z)$ up to approximately 0.129,
exceedingly larger than the measured value. This can be brought back inside
the experimental interval by the corrections from heavy thresholds,
provided there is a gap $\Lambda/M_c\approx 100$. This qualitative discussion
shows that in the present model gauge coupling unification can take place
with a compactification scale considerably smaller than the cut-off 
of the theory and, hopefully, smaller than the unification scale in conventional
GUTs.

\subsection{Estimate of $M_c$}

We have performed a numerical analysis in order to evaluate the 
compactification scale $M_c$. 
\begin{figure}[h!]
\begin{center}
\includegraphics[width=13.0cm]{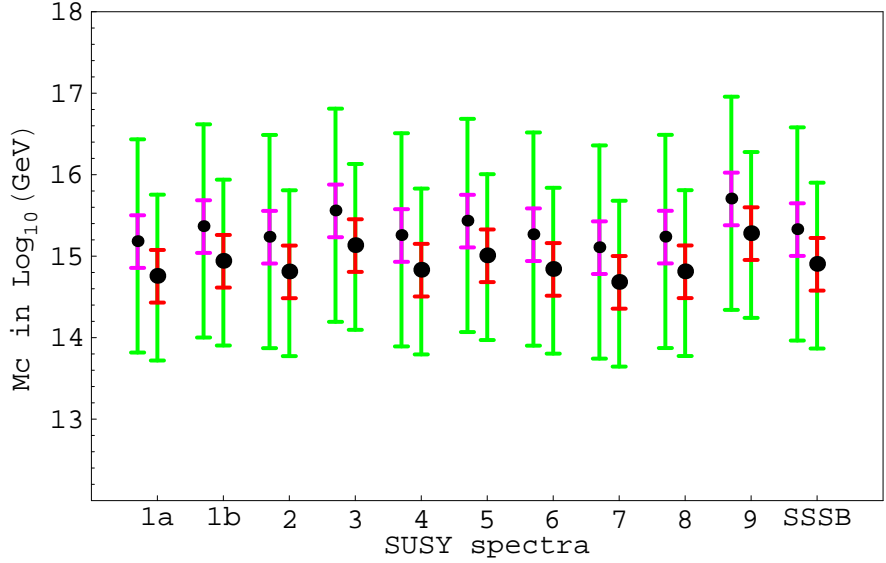}
\end{center}
\caption{Compactification scale $M_c$ versus SUSY spectrum.
For each SUSY spectrum, the bigger (smaller) circle on the right (left)
shows the result in the {\em HN} ({\em CPRT}) approach. The shorter
error bar represent the parametric error dominated by the experimental
uncertainty on $\alpha_3(m_Z)$, the wider bar includes the dominant source 
of error, the SU(5)-breaking brane terms.}
\label{mc}
\end{figure}
Inputs to our computations are the 
experimental results in eq. (\ref{inputrge}) and the spectra of supersymmetric
particles and extra Higgses given in section 2, tables 4 and 5. 
The brane contribution $\delta_i^{(b)}$ described in section 4.3 are taken 
as random variables in the interval of eq. (\ref{random}), where they 
are generated with a flat distribution.
Fig. 1 shows the compactification scale $M_c$ as a function of the spectrum
of supersymmetric particles. The heavy threshold corrections are evaluated
according to both the approaches {\em HN} (error bars on the right) and {\em CPRT} (error bars on the left) discussed in section 4.4. 
For each given SUSY spectrum there are 
two main sources of errors. One is the experimental error, approximately 
Gaussian, due to the uncertainties affecting the input parameters 
in eq. (\ref{inputrge}). This is largely dominated by the error
on $\alpha_3(m_Z)$ and is represented by the smallest error bar in the figures.
The other one is theoretical, associated to the unknown SU(5)-violating
kinetic terms at $y=\pi R/2$. This is a non-Gaussian error, since
we have no reason to prefer any values of the parameters $\delta_i^{(b)}$
in the interval (\ref{random}). The linear sum of these two errors
is described by the largest error bar in our figures.
The total error is therefore fully dominated by the theoretical one
and it heavily affects the prediction of the compactification scale, that
is predicted only up to an overall factor $10^{\pm 1}$. The prediction
of $M_c$ also depends on the spectrum of SUSY particles at the TeV scale.
The induced error on $M_c$ is however subdominant and the largest
difference between the values of $M_c$ obtained by varying the SUSY spectrum  
is by a factor of 4, at most.
Finally, also the treatment of the heavy thresholds gives rise to a 
theoretical uncertainty. The values of $M_c$ obtained by the {\em HN}
procedure are systematically smaller, by approximately a factor 2$\div$ 3, 
with respect to those given by the {\em CPRT} approach.
\begin{figure}[ht!]
\begin{center}
  \includegraphics[width=13.0cm]{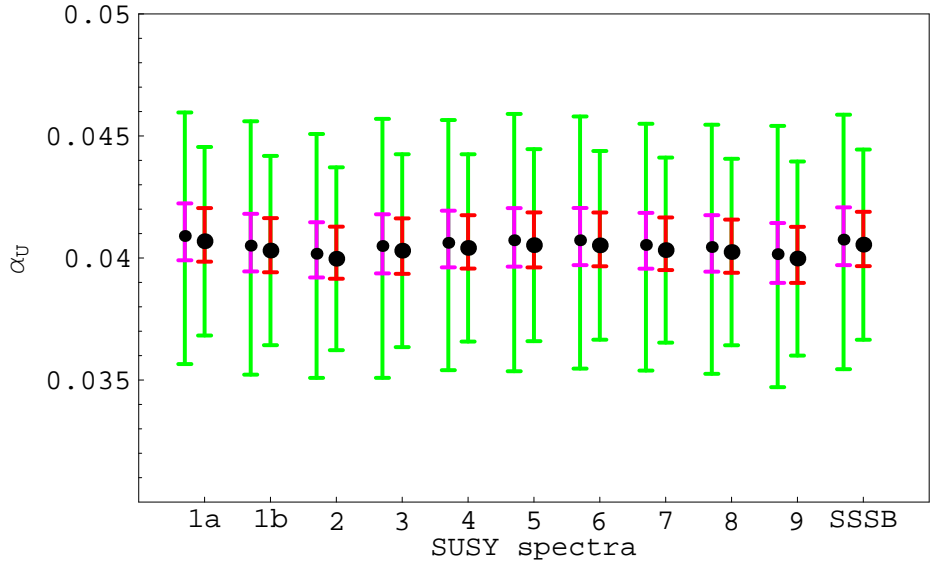}
\end{center}
\vskip 0.3cm
\caption{Gauge coupling $\alpha_U=g_U^2/(4\pi)$ at the compactification scale
versus SUSY spectrum. Error bars as in fig. \ref{mc}.}
\label{figalpha}     
\end{figure}
Despite the large overall uncertainty, from fig. 1 we see that on average the
compactification scale $M_c$ is smaller than the unification scale
of four-dimensional SUSY SU(5), 
$\approx 2\times 10^{16}$ GeV, by a factor of order 10.
The effective gauge boson mass, $M_X=2 M_c/\pi$, entering the dimension six 
operator of eqs. (\ref{lagpd0},\ref{lagpd}) is smaller than the 
corresponding mass 
in four-dimensional SUSY SU(5) by a factor $\approx 15$.
This produces an average enhancement by a factor 5$\times 10^4$ in the proton 
decay rate. Moreover, the key point in the present analysis is that
the theoretical uncertainties are completely dominated by 
the unknown SU(5) violating brane interactions and by the SUSY
spectrum. Given the present knowledge, 
we cannot prefer any brane interactions or any SUSY spectrum
among the various possibilities and the average prediction has not the meaning
of the most probable one. From this viewpoint values of $M_c$
as small as $10^{14}$ GeV are equally probable than $M_c\approx 10^{15}$ GeV,
and the enhancement of the proton decay rate can be as large as
5$\times 10^8$. 
\begin{figure}[h!]
\begin{center}
  \includegraphics[width=12.0cm]{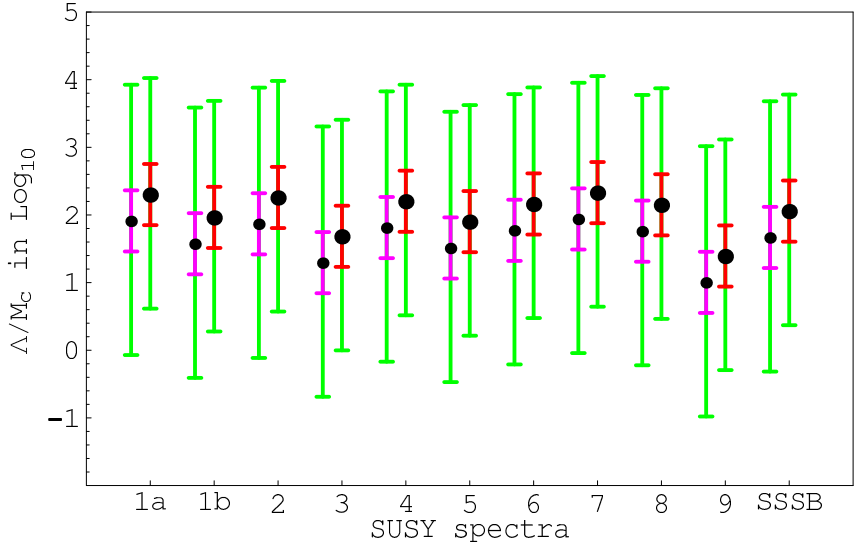}
\end{center}
\vskip 0.3cm
\caption{Ratio $\Lambda/M_c$ versus SUSY spectrum. Error bars as in fig.
\ref{mc}.}
\label{figratio}     
\end{figure}
Such an enhancement is sufficient to overcome the 
huge suppression factor coming from flavor mixing in options I and II, 
which in the most favorable case, as we can see from table 6, is 
$\vert\lambda^5\vert^2\approx 3\times 10^{-7}$. From 
these considerations we can already conclude that the proton decay rate
in the current model can be bigger by a factor $100$ compared
to the the rate evaluated in four-dimensional SUSY SU(5) through dimension
6 operators. As we will see
in the next section, we can obtain proton decay rates that are
quite interesting for the next generation of experiments.  
Alternatively when considering option 0, where no flavour suppressions
are present in the leading proton decay amplitudes, large values of 
$M_c$ such as $M_c\approx 10^{16}$ GeV allow to bring the
proton lifetime above the experimental lower limit in eq. (\ref{expbound}).
Fig. 2 shows the predicted gauge coupling $\alpha_U$ at the
compactification scale $M_c$. As we can see from eq. (\ref{lagpd}),
this is an input to our computation and we evaluate it as the average among $\alpha_i(M_c)$ $(i=1,2,3)$. 
We can see that $\alpha_U$
depends very mildly on the SUSY spectrum and on the treatment of 
heavy thresholds and is numerically close to 1/25, the value
in ordinary four-dimensional SUSY SU(5). The error bars in fig.
2 have the same meaning as in fig. 1. When computing proton decay
amplitudes what matters is the ratio $\alpha_U/M_c^2$, and we 
evaluate the errors on this quantity by fully accounting for 
correlations.

Finally, fig. 3 displays the dependence of the ratio $\Lambda/M_c$ 
on the SUSY spectrum, both for {\em HN} and {\em CPRT} cases.
We can see that the average value of $\Lambda/M_c$
is about $100$, what needed for the dominance of the bulk symmetric
SU(5) gauge coupling over the SU(5)-violating brane contribution in
eq. (\ref{Bandb}). The ratio $\Lambda/M_c$ is also directly related to the 
parameter $\epsilon$ of eq. (\ref{eps}), which controls the hierarchies
of Yukawa couplings. The range $\Lambda/M_c\approx 100$ is
consistent, within the present approximations, with 
$\epsilon\approx \lambda^2$.

%%%%%%%%%%%%%%%% 5. PROTON LIFETIME %%%%%%%%%%%%%%%%%%%%%%%%%%%%

\section{Proton lifetime}

In order to calculate the proton lifetime, we have to translate the operators in (\ref{lagpd0},\ref{lagpd},\ref{coefc}) 
at quark level to those at hadron level
using a perturbative chiral Lagrangian technique \cite{chl}. The aim is to evaluate the hadron matrix elements $\langle {\sl PS} \left|
{\cal O} \right| p \rangle$, which describe the transition from the proton to 
a pseudoscalar meson via the three-quark operator ${\cal O}$. 
\begin{figure}[h!]
\begin{center}
\includegraphics[width=12.0cm]{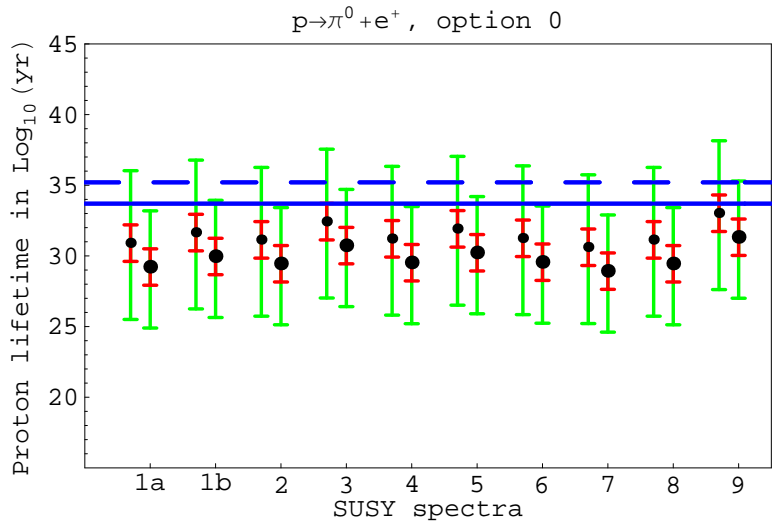}
\end{center}
\caption{Inverse decay rate of $p\to \pi^0 e^+$ 
versus the SUSY spectrum, in option 0.
The solid line is the current experimental lower bound, from ref. \cite{shio},
and the dashed line represents the aimed-for future sensitivity,
from ref. \cite{futsens}. Error bars as in fig. \ref{mc}.}
\label{fig0}
\end{figure}
The various matrix elements are calculated from the basic element
\be
  \alpha\, P_L\, u_p  = \epsilon_{\alpha\beta\gamma} \left\langle 0
    \left| \left( d_R^\alpha\, u_R^\beta \right) u_L^\gamma \right| p
  \right\rangle \,,
\label{hadmat}
\ee
where $u_p$ denotes the proton spinor and $\alpha$ 
is evaluated by means of lattice QCD simulations. Here we will adopt \cite{aoki}\footnote{The statistical error of the lattice estimate is less than 10\%,
but the systematic error is much larger. For instance, within the lattice
approach, ref. \cite{rcb} finds
$\left|\alpha\right|=0.007$ GeV$^3$. See also the discussion in ref.
\cite{raby}.}
\be
\left|\alpha\right|=0.015~~ {\rm GeV}^3~~~.  
\ee
In option 0, the dominant decay rate is given by:
\be
\Gamma(p\to e^+ \pi^0)  = 
\frac{\dd(m_p^2-m_{\pi^0}^2)^2}{\dd 64\pi m_p^3 f_\pi^2}\, 
{\alpha^2}\, A^2_SA^2_L\; \frac{\dd g_U^4}{\dd M^4_X} 
\left(1+D+F\right)^2\, \left[\left(1+\vert V_{ud}\vert^2\right)+1\right]~~~.
\ee
\begin{figure}[h!]
\begin{center}
\includegraphics[width=12.0cm]{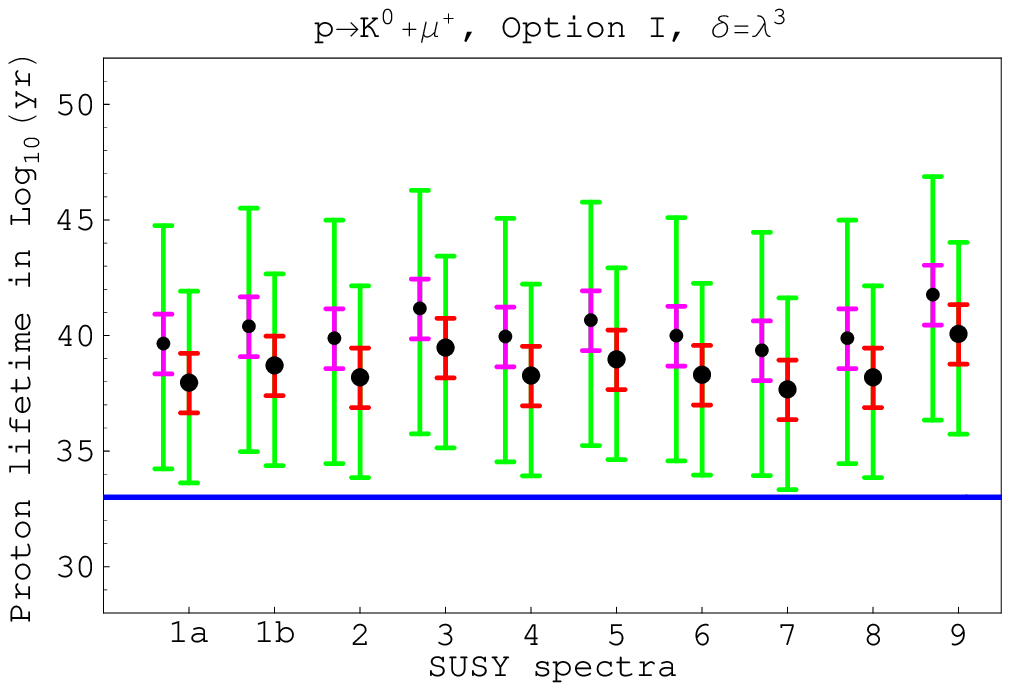}
\includegraphics[width=12.0cm]{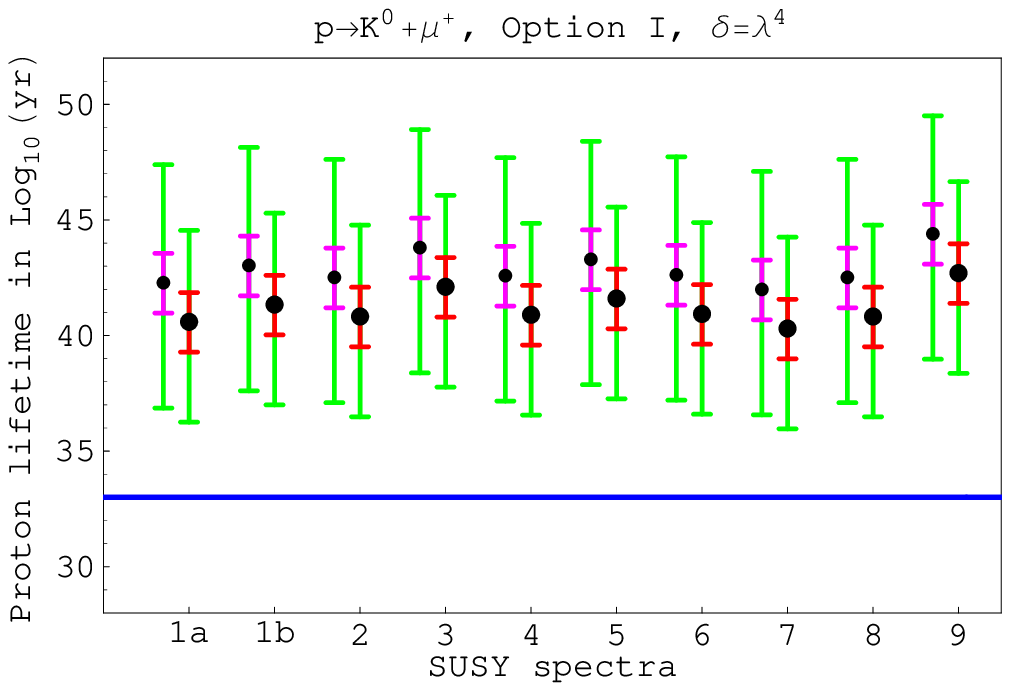}
\end{center}
\caption{Inverse decay rate of $p\to K^0 \mu^+$ versus the SUSY spectrum, 
in option I, 
for $\delta=\lambda^3$ (upper panel) and $\delta=\lambda^4$ (lower panel).
The solid line is the current experimental lower bound, from ref. \cite{shio}.
Error bars as in fig. \ref{mc}.}
\label{fig1}
\end{figure}
Here
$m_p$, $m_\pi$ denote the proton and pion masses
respectively, and $f_\pi=130$ MeV is the pion decay constant;
$D$ and $F$ are the symmetric and antisymmetric SU(3)
reduced matrix elements for the axial-vector current and recent
hyperon decays measurements \cite{cabibbo03} give $D=0.80$ and
$F=0.46$. The effect of lepton masses is neglected. The short- and
long-distance renormalization factors $A_S$,
$A_L$ encode the evolution from the GUT scale to the
SUSY-breaking scale and the evolution from the SUSY-breaking scale
to 1\,GeV \cite{asl}. We use:
\be
A_L=
\left(\frac{\alpha_3(\mu_{had})}{\alpha_3(m_c)}\right)^{\frac{2}{9}}
\left(\frac{\alpha_3(m_c)}{\alpha_3(m_b)}\right)^{\frac{6}{25}}
\left(\frac{\alpha_3(m_b)}{\alpha_3(m_Z)}\right)^{\frac{6}{23}}
\approx 1.43
\ee
with $\mu_{had}=1$ GeV and
\be
A_S=
\left(\frac{\alpha_1(m_Z)}{\alpha_U}\right)^{\frac{23}{30 b_1}}
\left(\frac{\alpha_2(m_Z)}{\alpha_U}\right)^{\frac{3}{2 b_2}}
\left(\frac{\alpha_3(m_Z)}{\alpha_U}\right)^{\frac{4}{3 b_3}}
\approx 2.37
\ee
where the U(1) contribution is only approximate.
\begin{figure}[h!]
\begin{center}
\includegraphics[width=12.0cm]{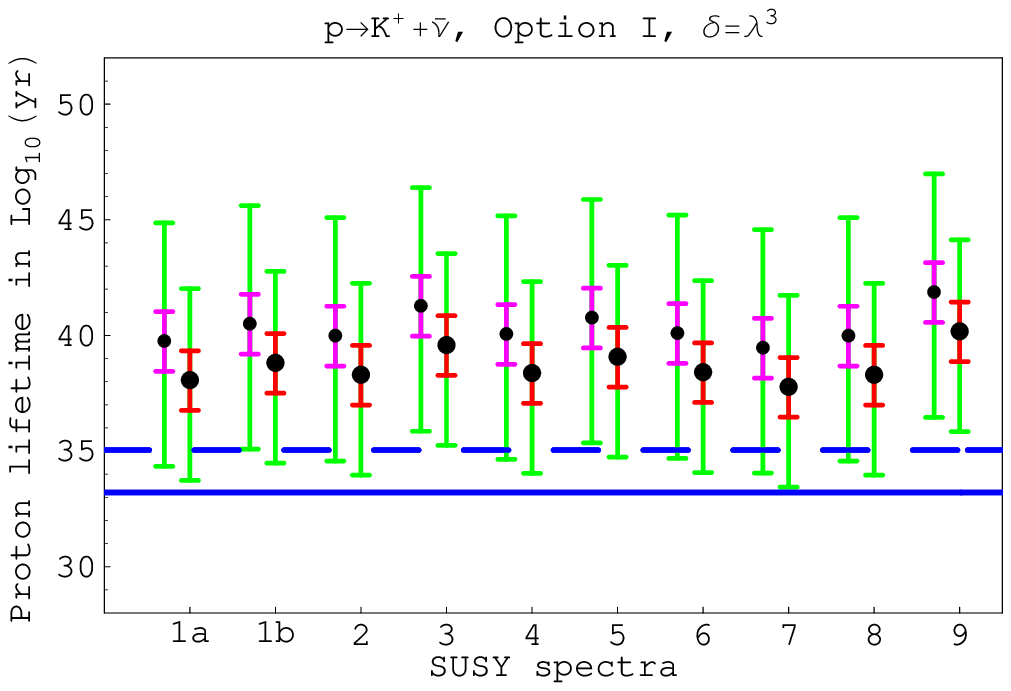}
\includegraphics[width=12.0cm]{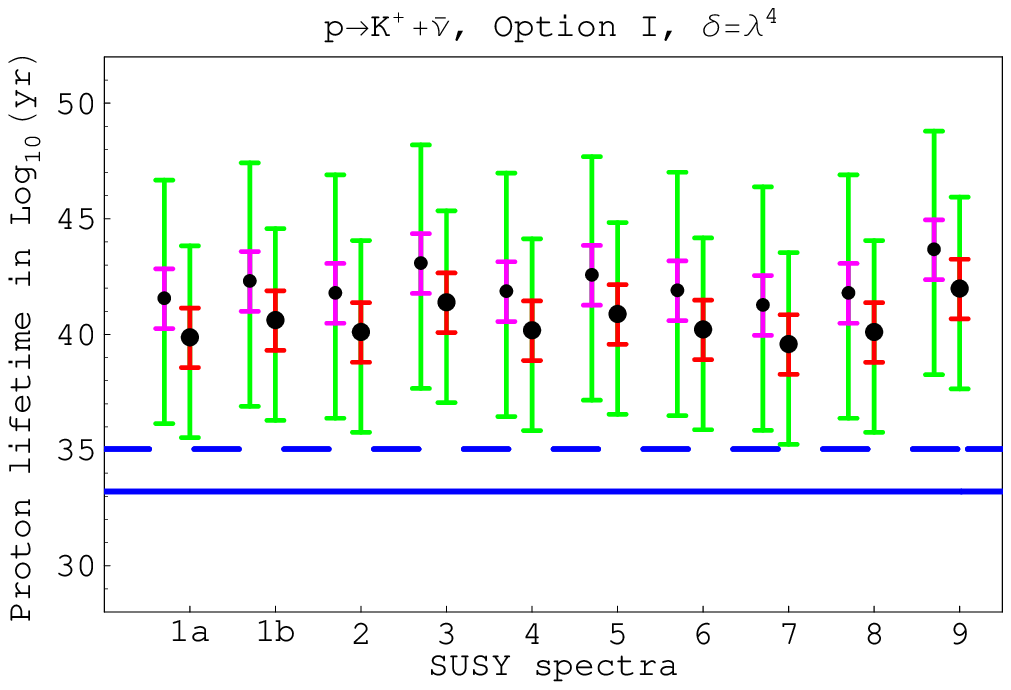}
\end{center}
\caption{Inverse decay rate of $p\to K^+ \nu$ (summed over all neutrino channels) versus the SUSY spectrum, 
in option I, 
for $\delta=\lambda^3$ (upper panel) and $\delta=\lambda^4$ (lower panel).
The solid line is the current experimental lower bound, from ref. \cite{shio},
and the dashed line represents the aimed-for future sensitivity,
from ref. \cite{futsens}. Error bars as in fig. \ref{mc}.}
\label{fig2}
\end{figure}
In fig. \ref{fig0} we see the predicted inverse decay rate
as a function of the SUSY spectrum.
A general comment, that applies to all the results concerning the rates
in options 0, I and II,
is that there is a huge theoretical uncertainty that spreads over
many order of magnitudes. By far, the main source of this uncertainty 
is the compactification scale $M_c$, which is only know up
to about two order of magnitudes, due to the unknown SU(5)-breaking
brane contribution. Since the proton lifetime scales as $M_c^4$,
this corresponds to an uncertainty of more than eight order of magnitudes
on the inverse rates. In this enormous range the probability is however
almost uniform. From fig. \ref{fig0} we see that most of the parameter space
of the model have already been excluded by the existing experimental bound.
Though the model is not entirely ruled out and the allowed region 
in parameter space will be almost completely explored by the next 
generation of experiments.
\begin{figure}[h!]
\begin{center}
\includegraphics[width=12.0cm]{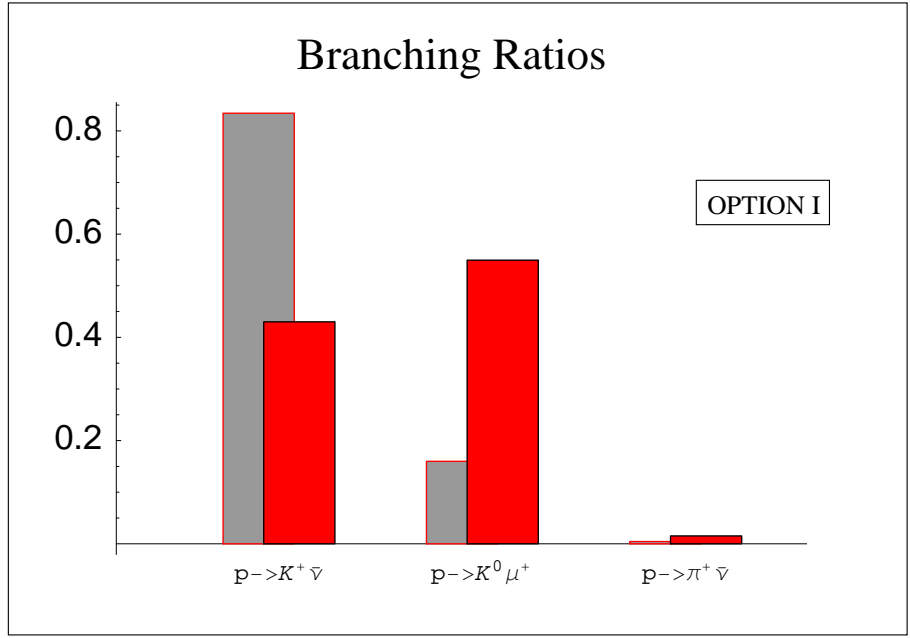}
\end{center}
\caption{Branching ratios for option I. Dark-red (light-grey) histograms 
refer to $\delta=\lambda^3$ ($\delta=\lambda^4$).}
\label{brI}
\end{figure}
The decay rates of option I and II for the different channels are given by: 
\bea
\Gamma(p\to e_j^+ \pi^0) & = &
\frac{\dd(m_p^2-m_{\pi^0}^2)^2}{\dd 64\pi m_p^3 f_\pi^2}\, 
{\alpha^2}\, A^2_SA^2_L\; \frac{\dd g_U^4}{\dd M^4_X} 
\left(1+D+F\right)^2\, \left[\vert C^1_{j1} \vert^2 + \vert C^2_{j1} \vert^2 \right]
\nn\\
\Gamma(p\to e_j^+ K^0) & = &
\frac{\dd(m_p^2-m_{K^0}^2)^2}{\dd 32\pi m_p^3 f_\pi^2}\, 
{\alpha^2}\,  A^2_SA^2_L\; \frac{\dd g_U^4}{\dd M^4_X} 
\left(1+\dd\frac{m_p}{m_B}(D-F) \right)^2\,  
\left[\vert C^1_{j2} \vert^2 + \vert C^2_{j2} \vert^2 \right]
\nn\\
\Gamma(p\to \bar\nu_j \pi^+) & = &
\dd\frac{(m_p^2-m_{\pi^\pm}^2)^2}{32\pi m_p^3 f_\pi^2}\,
{\alpha^2}\, A^2_S A^2_L\; \dd\frac{g_U^4}{M^4_X} 
\left( 1+D+F \right)^2\,  \vert C^3_{11j} \vert^2
\nn\\
\Gamma(p\to \bar\nu_j K^+) & = &
\dd\frac{(m_p^2-m_{K^\pm}^2)^2}{32\pi m_p^3 f_\pi^2}\, 
{\alpha^2} A^2_S A^2_L\; \dd\frac{g_U^4}{M^4_X} \times
\nn\\
& &\times
\left\vert
\dd\frac{2}{3} \dd\frac{m_p}{m_B} D~ C^3_{12j} \!+\! 
\left(1+\dd\frac{m_p}{3 m_B} (D+3F)\right) C^3_{21j} 
\right\vert^2~~~.
\label{rates}
\eea
Here $m_K$ denote the kaon mass and $m_B=1.15$ GeV is an average 
baryon mass according to contributions from diagrams with virtual 
$\Sigma$ and $\Lambda$. In our numerical 
estimates the interference in the expression of the rate for $K^+ {\bar\nu}$
is constructive.
\begin{figure}[h!]
\begin{center}
\includegraphics[width=12.0cm]{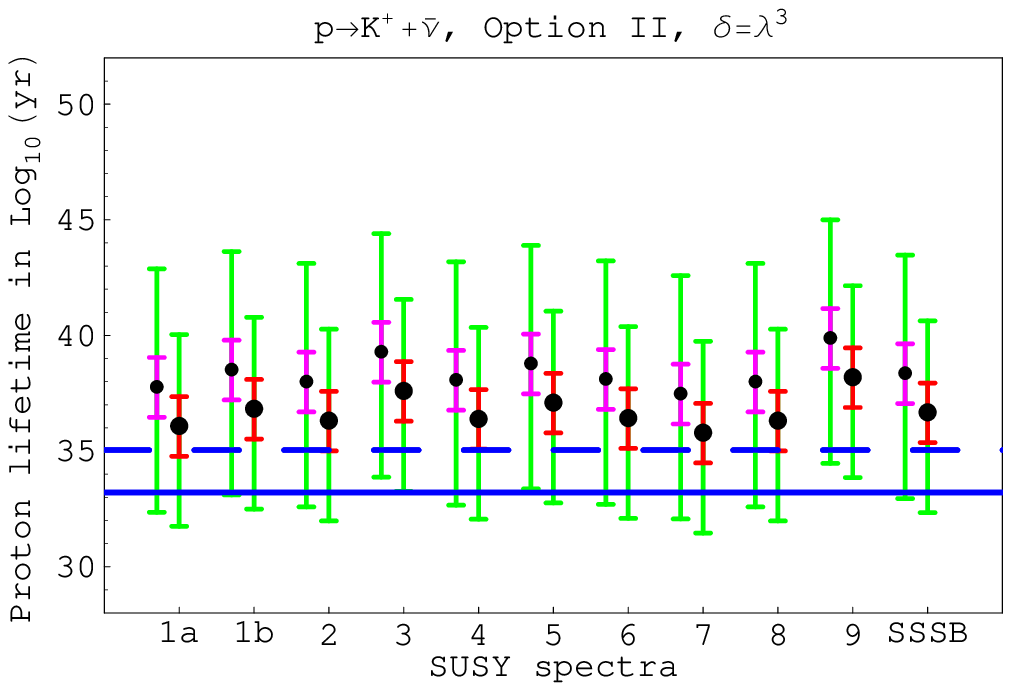}
\includegraphics[width=12.0cm]{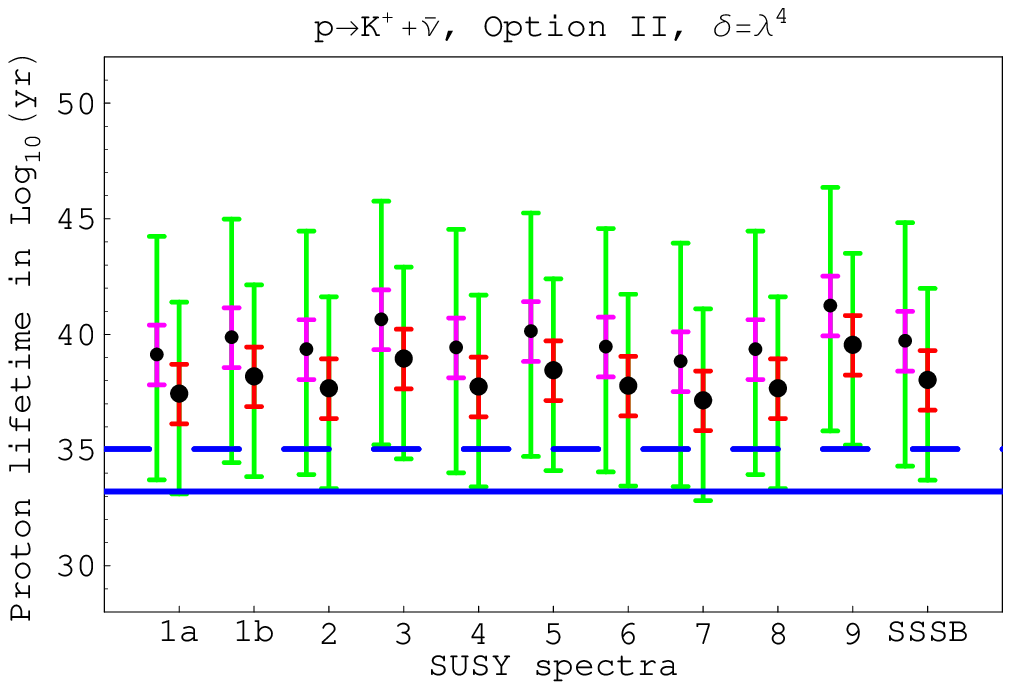}
\end{center}
\caption{Inverse decay rate of $p\to K^+ {\bar \nu}$ (summed over all neutrino channels) versus the SUSY spectrum, 
in option II, 
for $\delta=\lambda^3$ (upper panel) and $\delta=\lambda^4$ (lower panel).
The solid line is the current experimental lower bound, from ref. \cite{shio},
and the dashed line represents the aimed-for future sensitivity,
from ref. \cite{futsens}. Error bars as in fig. \ref{mc}.}
\label{fig3}
\end{figure}
The dominant proton decay rates are shown in fig. (\ref{fig1},\ref{fig2})
and (\ref{fig3}-\ref{fig5}).
The results strongly depend on the location of $F_1$, either in the bulk
(option I) or on the brane at $y=0$ (option II). 
As we can see from Table 4, if $F_1$ is a bulk field, then the dominant
four-fermion operators are $uus\mu$ and $uds\nu_{\mu,\tau}$ and,
consequently, the preferred proton decay channels are $K^0 \mu^+$ and
$K^+ \nu$, whose rates are displayed in fig. (\ref{fig1},\ref{fig2}). 
We see from fig. (\ref{fig1},\ref{fig2}) 
that the possibilities of testing proton decay in option I are quite
remote. If $\delta$, which parametrizes our ignorance about the mixing
angles, is equal to $\lambda^4$, the inverse rates are always larger than
10$^{35}$ yr, too long to be observed in future planned facilities.
When $\delta=\lambda^3$, only in some specific case the inverse rates can reach
10$^{34}$ yr, depending on the SUSY spectrum and on 
a favourable combination of the brane corrections $\delta_i^{(b)}$.
\begin{figure}[h!]
\begin{center}
\includegraphics[width=12.0cm]{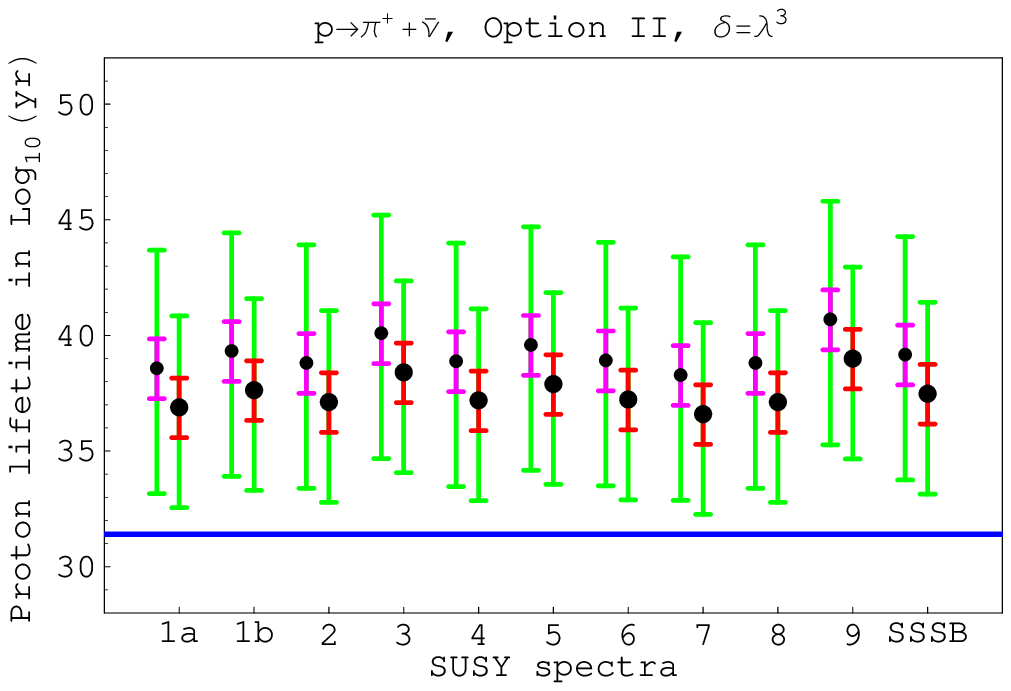}
\includegraphics[width=12.0cm]{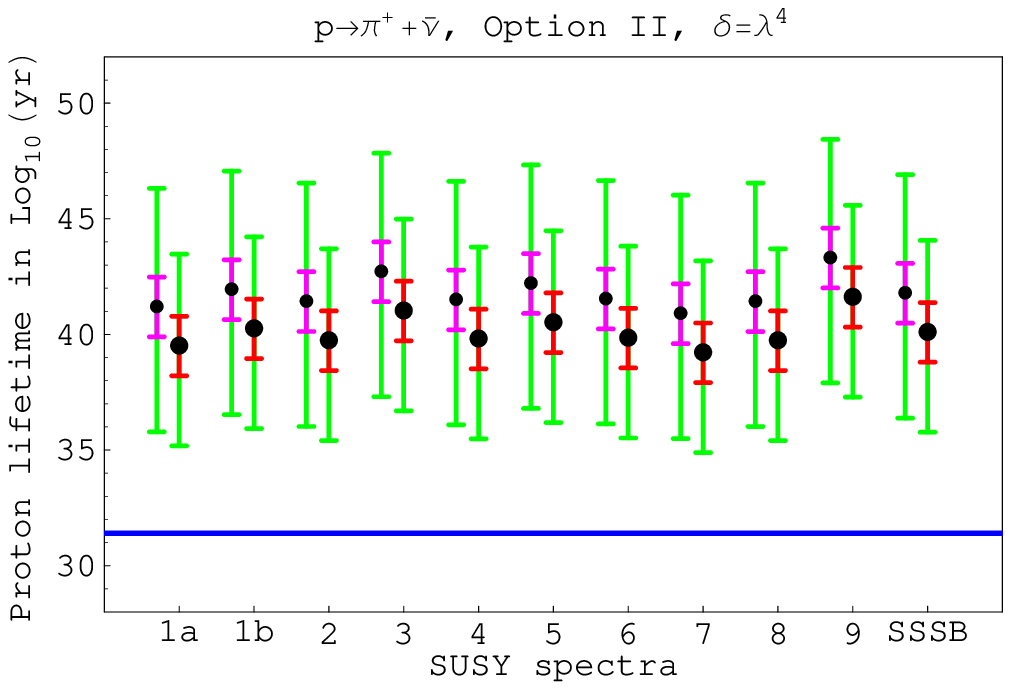}
\end{center}
\caption{Inverse decay rate of $p\to \pi^+ {\bar \nu}$ (summed over all neutrino channels) versus the SUSY spectrum, 
in option II, 
for $\delta=\lambda^3$ (upper panel) and $\delta=\lambda^4$ (lower panel).
The solid line is the current experimental lower bound, from ref. \cite{shio}. 
Error bars as in fig. \ref{mc}.}
\label{fig4}
\end{figure}
While the absolute rates are affected by a very large theoretical uncertainty,
the predictions about the branching ratios are considerably more precise.
The dependence on $M_c$ cancels out in the branching ratios and therefore
the relative decay rates do not depend neither on the unknown SU(5)-breaking
brane terms, nor on the SUSY spectrum. The uncertainty is dominated by
the mixing matrices and is parametrized in our discussion by $\delta$,
which we let vary between $\lambda^3$ and $\lambda^4$. In fig. \ref{brI}
we see the BRs of the dominant channels in option I, $K^+ {\bar \nu}$ and
$K^0 \mu^+$, which are comparable
within the estimated uncertainty.
\begin{figure}[h!]
\begin{center}
\includegraphics[width=12.0cm]{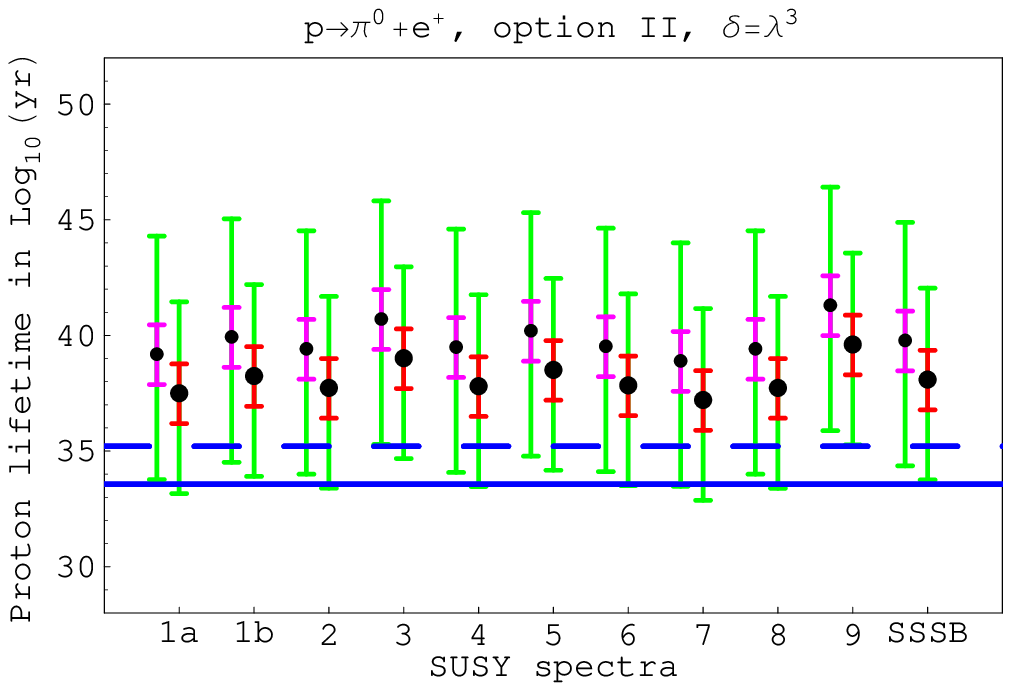}
\includegraphics[width=12.0cm]{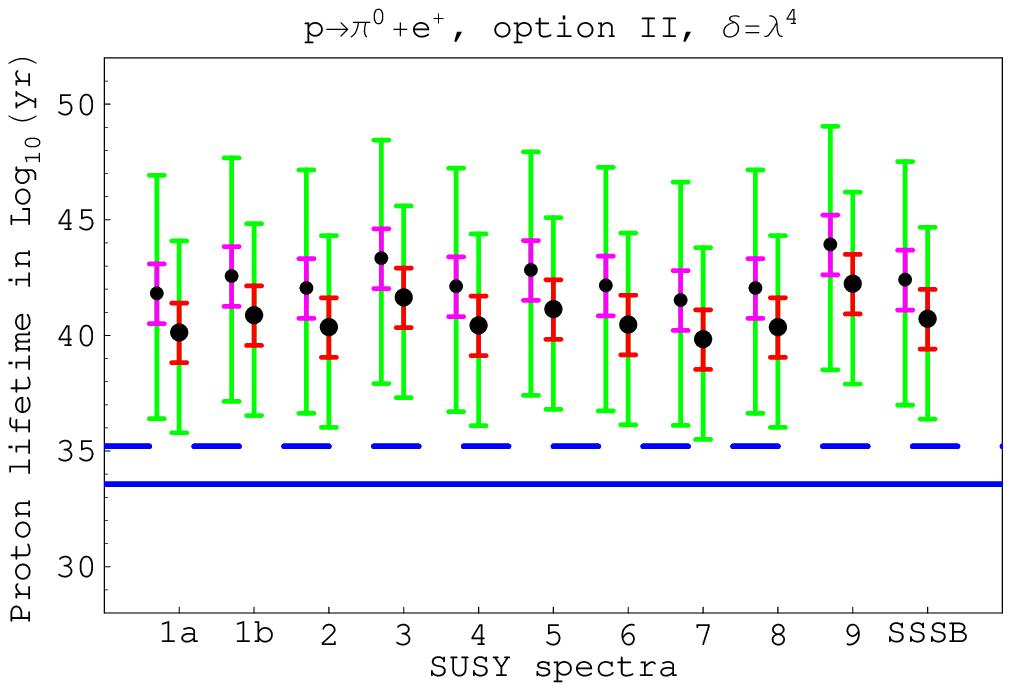}
\end{center}
\caption{Inverse decay rate of $p\to \pi^0 e^+$ (summed over all neutrino channels) versus the SUSY spectrum, 
in option II, 
for $\delta=\lambda^3$ (upper panel) and $\delta=\lambda^4$ (lower panel).
The solid line is the current experimental lower bound, from ref. \cite{shio},
and the dashed line represents the aimed-for future sensitivity,
from ref. \cite{futsens}. Error bars as in fig. \ref{mc}.}
\label{fig5}
\end{figure}
Substantially different is the option II, where $F_1$ is a brane field
like $F_{2,3}$. Here the preferred decay channel is $K^+ {\bar \nu}$,
as we can guess from Table 6, showing that this final state has the smallest
flavour suppression. As we can see from fig. 8, a small portion of the 
parameter space of the model has already been excluded by the 
current experimental bound
and, when $\delta=\lambda^3$, there will be good chances
for the future facilities to discover proton decay in this channel,
for almost all possible type of SUSY spectra considered in our analysis.
Also if $\delta=\lambda^4$ and, consequently,
the rates are more suppressed, there are several SUSY spectra which would
allow a lifetime below $10^{35}$ yr, at least for some combination 
of the brane kinetic parameters $\delta_i^{(b)}$. 

The rate for $\pi^+ {\bar \nu}$ is down by approximately a factor $\lambda^2$,
compared with the dominant one. Only for $\delta=\lambda^3$, for some
specific SUSY spectra and in a favourable range of $\delta^{(b)}_i$,
the predicted inverse rates are below $10^{34}$ yr, which would however 
represent an improvement over the current limit by more than two orders 
of magnitudes.

The prediction for the decay channels containing a charged lepton
are similar to those for $\pi^+ {\bar \nu}$, but the expected future
experimental sensitivity is quite better, especially in the $\pi^0 e^+$
mode, as shown in fig. \ref{fig5}. A modest portion of the parameter
space, requiring $\delta$ close to $\lambda^3$, leads to inverse rates
that are within the reach of future facilities.
\begin{figure}[h!]
\begin{center}
\includegraphics[width=12.0cm]{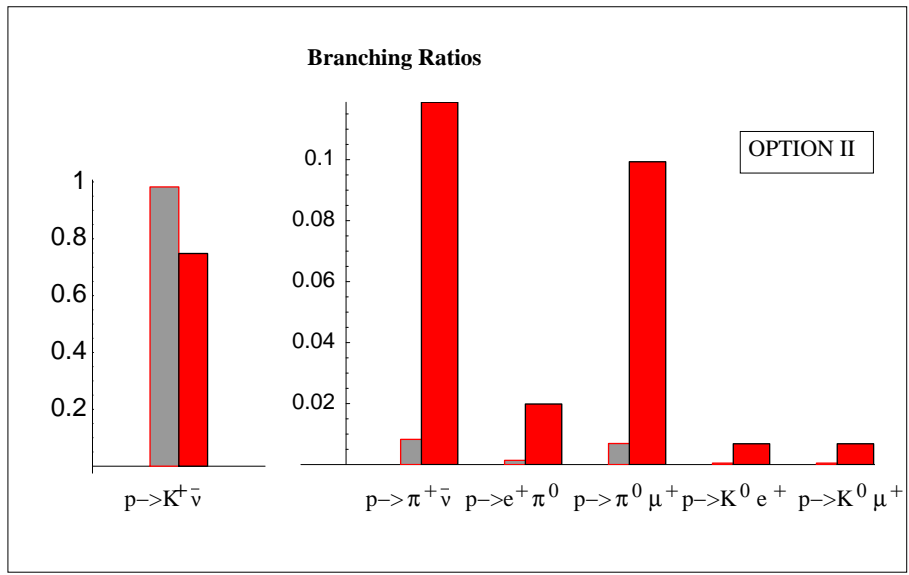}
\end{center}
\caption{Branching ratios for option II. Dark-red (light-grey) histograms 
refer to $\delta=\lambda^3$ ($\delta=\lambda^4$).}
\label{brII}
\end{figure}
The branching ratios for option II are displayed in fig. \ref{brII}.
We can see the dominance of $K^+ {\bar \nu}$, followed by the 
$\pi^+ {\bar \nu}$ and 
$\pi^0 \mu^+$ channels. The other
channels with a charged lepton, such as $\pi^0 e^+$, $K^0 e^+$ 
and $K^0 \mu^+$ are comparable, but up to now future experimental prospects
have been clearly worked out only for the $\pi^0 e^+$ mode. 

%%%%%%%%%%%%%%%%%%%%%%% CONCLUSION %%%%%%%%%%%%%%%%%%%%%%%%%%%%%%

\section{Conclusion}
The decisive test of grand unification theories is proton decay. Minimal SUSY
GUTs are strongly disfavoured today by the stringent experimental bounds on 
proton lifetime. Moreover such minimal realizations of the grand unification 
idea are also plagued by serious theoretical problems, first of all the 
doublet-triplet splitting problem, requiring a tuning by fourteen orders
of magnitude. An elegant solution of the doublet-triplet splitting problem
is offered by SUSY GUTs where the GUT symmetry is broken by the 
compactification of an extra dimension. There is a whole class of
such GUTs where the leading baryon-violating operators arise from
gauge vector boson exchange, have dimension six and scale as the
inverse of the compactification mass $M_c$ squared.

We have evaluated $M_c$ form a next-to-leading analysis
of gauge coupling unification including SUSY and GUT threshold
corrections, two-loop running and SU(5)-breaking boundary terms.
We have carefully estimated the uncertainties due to
experimental errors and poor theoretical knowledge, such as 
the ignorance about the SUSY breaking spectrum and about the
details of the SU(5)-breaking brane dynamics. In view of the existing
intrinsic theoretical uncertainties we have found a wide range
of acceptable values for $M_c$, going from approximately $10^{14}$ GeV
to more than $10^{16}$ GeV.

The consequences on proton decay strongly depend on the features
of the flavour sector, which can be considerably different from the standard, 
four-dimensional one. In the simplest case all matter fields have
the same location in the extra space and proton decay proceeds mainly
via the $\pi^0 e^+$ channel, without any flavour suppression.
Such a scenario is strongly disfavoured by now, since $M_c$ is on average
much smaller than the four-dimensional unification scale $\approx 2\times
10^{16}$ GeV. However, due to the large uncertainties on $M_c$, it is
not yet ruled out and future facilities can essentially complete the 
test of this model.

When matter fields have no identical location in the extra dimension, 
fermion mass hierarchies and mixing angles are more naturally described.
In our analysis we have considered two options, corresponding to
``semianarchy'' (option I) and ``anarchy'' (option II) in the neutrino sector.
In option I the suppression coming from flavour mixing is too strong and 
proton decay is almost beyond the possibilities 
of the next generation of experiments.
In option II the flavour suppression of the $K^+ {\bar \nu}$ channel can be
compensated by the allowed smallness of $M_c$ resulting in a proton lifetime 
well within the reach of future facilities.

\vspace*{1.0cm}
{\bf Acknowledgements}
We thank Luigi Pilo for useful discussions. This project is partially
supported by the European Program MRTN-CT-2004-503369.
\vfill

\end{document}